\documentclass[11pt]{article}
\usepackage{putex}
\usepackage{graphicx}

\begin{document}
\preprint{PUPT-2475}

\title{Holographic Fermi surfaces at finite temperature in top-down constructions}
\authors{Charles Cosnier-Horeau$^\ENS$ and Steven S. Gubser$^\PU$}
\institution{ENS}{${}^1$D\'epartement de Physique, \'Ecole Normale Sup\'erieure, 45 rue d'Ulm, 75005 Paris, France}
\institution{PU}{${}^2$Joseph Henry Laboratories, Princeton University, Princeton, NJ 08544, USA}
\abstract{
We calculate the two-point Green's functions of operators dual to fermions of maximal gauged supergravity in four and five dimensions, in finite temperature backgrounds with finite charge density.  The numerical method used in these calculations is based on differential equations for bilinears of the supergravity fermions rather than the equations of motion for the fermions themselves.  The backgrounds we study have vanishing entropy density in appropriate extremal limits.  Holographic Fermi surfaces are observed when the scalar field participating in the dual field theory operator has an expectation value, which makes sense from the point of view that the quasi-particles near the Fermi surfaces observed carry non-singlet gauge quantum numbers in the dual field theory.}
\date{November 2014}

\maketitle

\tableofcontents

\section{Introduction}

The original approach to studying Fermi surfaces in holography was to solve some variant of the Dirac equation in an anti-de Sitter (AdS) Reissner-Nordstrom (RN) background: see for example \cite{Lee:2008xf,Liu:2009dm,Cubrovic:2009ye}.  A two-point Green's function can be extracted from solutions to the Dirac equation, and a singularity in that Green's function at $\omega=0$ and $k=k_F$, corresponding to a normalizable fermion mode, signals the existence of a Fermi surface.  This bottom-up methodology has the advantage of simplicity and flexibility, and a range of interesting behaviors have been uncovered.  

A top-down approach has been pursued in \cite{DeWolfe:2011aa,DeWolfe:2012uv,DeWolfe:2013uba,DeWolfe:2014ifa} (see also earlier work based on probe branes \cite{Ammon:2010pg} and negative results \cite{Gauntlett:2011mf,Belliard:2011qq,Gauntlett:2011wm} for gravitinos) based on maximal gauged supergravity in four and five dimensions.  While more technically involved than the bottom-up approach, top-down has the advantage that the dual theories are known, and there is a simple way to suppress the zero-point entropy that plagues interpretations of the AdSRN results, namely to consider backgrounds in which one or more of the several commuting $U(1)$ charges vanishes.  Holographic Fermi surfaces were found in both the $AdS_4$ and $AdS_5$ top-down constructions.  Indeed, it is possible to read off from \cite{DeWolfe:2012uv,DeWolfe:2014ifa} detailed answers to the question of whether specific fermionic operators exhibit Fermi surface singularities, in black holes backgrounds at zero temperature and non-zero chemical potential for all of the commuting $U(1)$ gauge charges.  Some results for the zero-entropy cases, where one or more $U(1)$ charge vanishes, are available from \cite{DeWolfe:2012uv,DeWolfe:2013uba,DeWolfe:2014ifa}, and these results match well with calculations at low but non-zero temperature to be presented in the current work.

The key question left outstanding is what field theory mechanisms are at work in the dual field theories driving the particular pattern of Fermi surfaces observed from the top-down supergravity calculations.  Also, the calculations done so far in the top-down formalism focus on identifying fermionic zero modes at zero temperature, and it is worthwhile to extend these analyses to computations of the full Green's functions.  In this paper, we aim to fill these gaps, keeping temperature finite but focusing on the cases where entropy vanishes in an extremal limit.  Along the way, in section~\ref{METHOD}, we will introduce an efficient numerical method (more efficient than the ones commonly found in the literature) for extracting the fermionic Green's functions.  In section~\ref{NUMERICS} we will present a survey of numerical results obtained using this method, and we will formulate a heuristic ``boson rule'' determining when holographic Fermi surfaces appear for backgrounds with no zero-point entropy.  Briefly, the boson rule says that there is a Fermi surface when the color-singlet fermionic operator in the field theory involves a colored boson which has a condensate.  Our numerical exploration also supports a ``fermion rule,'' which states that the value of $k_F$ is suppressed, though it may not vanish, when the color-singlet fermionic operator involves a colored fermion which is neutral under the $U(1)$ charge carried by the black hole.  Both of these phenomenological rules are readily intelligible under the ``gaugino interpretation'' of \cite{DeWolfe:2011aa}, in which holographic Fermi surfaces are held to be Fermi surfaces of colored fermions in the dual gauge theory.

\section{A Green's function method based on bulk currents}
\label{METHOD}

\subsection{The fermionic equations of motion}

Our results come from solving equations of the form
 \eqn{DiracMostlyMinus}{
\left[ i \gamma^\mu \nabla_\mu -  g \left( m_1 e^{-\phi \over \sqrt{6}} + m_2 e^{2\phi \over \sqrt{6}} \right) +g q_1 \gamma^\mu a_\mu +gq_2 \gamma^\mu A_\mu + i p_1 e^{-2\phi\over \sqrt{6}} f_{\mu\nu} \gamma^{\mu\nu} + i p_2 e^{\phi\over \sqrt{6}} F_{\mu\nu} \gamma^{\mu\nu} \right] \chi = 0
 }
in five dimensions, and
 \eqn{DiracFour}{
  \left[ i \gamma^\mu \nabla_\mu + {1 \over 4L} \sum_{i=a,b,c,d} m_i e^{\lambda_i/2} + 
    {1 \over 4L} \gamma^\mu \sum_{i=a,b,c,d} q_i A^i_\mu + {i \over 8}
     \sigma^{\mu\nu} \sum_{i=a,b,c,d} \left( p_i e^{-\lambda_i/2} F_{\mu\nu}^i \right) \right] \chi = 0
 }
in four dimensions.  Points to note are:
 \begin{itemize}
  \item \eno{DiracMostlyMinus} is equation (80) of \cite{DeWolfe:2012uv}.
  \item \eno{DiracFour} is equation (40) of \cite{DeWolfe:2014ifa}.
  \item $\nabla_\mu$ includes the spin connection but not the gauge connections.
  \item $F_{\mu\nu}$ and $f_{\mu\nu}$, with gauge potentials $A_\mu$ and $a_\mu$, correspond to a particular combination of the three $U(1)$ factors which form a Cartan subalgebra of $SO(6)$, which is the gauge group of five-dimensional maximal gauged supergravity.  Namely, $f_{\mu\nu}$ pertains to what we will consider the first $U(1)$, while $F_{\mu\nu}$ is the diagonal combination of the second and third.  The black hole backgrounds we will study carry electric charge under either $F_{\mu\nu}$ or $f_{\mu\nu}$, but not both.  The parameter $g$ is the $SO(6)$ gauge coupling, and $g = 2/L$ where $L$ is the radius of $AdS_5$.
  \item $F_{\mu\nu}^i$ denote gauge fields, with gauge potentials $A_\mu^i$, corresponding to the four commuting $U(1)$ factors which form a Cartan subalgebra of $SO(8)$, which is the gauge group of four-dimensional maximal gauged supergravity.  The black hole backgrounds we will study carry electric charge under some but not all of these $U(1)$s.
  \item $\phi$ in \eno{DiracMostlyMinus} is scalar in the ${\bf 20}'$ of $SO(6)$.  It is electrically neutral with respect to $A_\mu$ and $a_\mu$.
  \item The $\lambda_i$ in \eno{DiracFour} are scalars in the ${\bf 35_v}$ of $SO(8)$ and are subject to the constraint $\sum_{i=a,b,c,d} \lambda_i = 0$.  They are electrically neutral with respect to the $A_\mu^i$.
  \item The parameters $(m_1,m_2,q_1,q_2,p_1,p_2)$ in the five-dimensional case, and the parameters $(m_i,q_i,$ $p_i)$ in the four-dimensional case, are entirely determined by the structure of maximal gauged supergravity.  The cases we study are listed in tables~\ref{FiveDTable} and~\ref{FourDTable}.
  \item $\chi$ denotes a four-component spinor.  In five-dimensional supergravity, it comes from the part of the ${\bf 48}$ of the local symmetry group $USp(8)$ which does not mix with the gravitinos in the backgrounds of interest.  In four-dimensional supergravity, it comes from the part of the ${\bf 56}$ of the local symmetry group $SU(8)$ which does not mix with the gravitinos.  The operators dual to this type of fermion all take the form $\tr \lambda X$, which have dimension $5/2$ in the $AdS_5$ case and $3/2$ in the $AdS_4$ case.
 \end{itemize}
Readers wishing to skip over supergravity calculations and see the results for the Green's functions of the operators dual to $\chi$ may at this point go straight to section~\ref{NUMERICS}.

All backgrounds under consideration in this paper are planar, electrically charged black holes which are asymptotically $AdS_5$ or $AdS_4$.  The metrics take the form
 \eqn{BackgroundMetric}{
  ds^2 = e^{2A} \left( -h dt^2 + d\vec{x}^2 \right) + {e^{2B} \over h} dr^2 \,,
 }
where $A$, $B$, and $h$ are functions of $r$, and $\vec{x} \in {\bf R}^3$ for $AdS_5$ and ${\bf R}^2$ for $AdS_4$.  Pure $AdS_5$ or $AdS_4$ would be recovered by setting $A = -B = \log {r \over L}$ and $h=1$.  We focus on backgrounds with regular horizons, in which $h$ decreases from $1$ at the boundary to $0$ at a radius $r=r_H$.  In our coordinate systems, $r_H \to 0$ is the extremal limit.

Explicit expressions for $A$, $B$, $h$, $\Phi_1 \equiv a_t$, $\Phi_2 \equiv A_t$, and $\phi$ can be found in equation (3) of \cite{DeWolfe:2012uv}.  In equations (4)-(5) of the same paper, expressions can be found for the temperature $T$; the chemical potentials $\mu_1$ and $\mu_2$, relating, respectively, to $a_\mu$ and $A_\mu$; the entropy density $s$; and the charge densities $\rho_1$ and $\rho_2$.  All quantities can be expressed in terms of charge parameters $Q_1$ and $Q_2$ and the horizon radius $r_H$, all of which have dimensions of length.  We will set $L=1$ in both five and four dimensions from here on.

Expressions for the metric, the electrostatic potentials $\Phi_i$, and the scalars $\lambda_i$ can be found in (18)-(21) of \cite{DeWolfe:2014ifa} (note the definitions (14) in that paper of the $\lambda_i$ and the sign convention $\eta_i=1$ for the backgrounds we study in the current work).  In (25)-(28) of the same paper, expressions are given for the temperature $T$; the four chemical potentials $\mu_i$, related to the gauge fields $A_\mu^i$; the entropy density $s$; and the charge densities $\rho_i$.  All quantities can be expressed in terms of four charge parameters $Q_i$ and the horizon radius $r_H$.

Because there is translation invariance in the $t$ and $\vec{x}$ directions, we require $\chi \propto e^{-i \omega t + i k x^1}$, where $x^1$ is one of the spatial directions.  Green's functions in the dual field theory will be functions of $\omega$ and $k$.  More explicitly, with a choice of Clifford basis as in \cite{DeWolfe:2012uv,DeWolfe:2014ifa}, we use an ansatz
 \eqn{ChiAnsatz}{
  \chi = {e^{-i\omega t + ikx^1} \over \sqrt[4]{\det g_{mn}}}
    \begin{pmatrix} \psi_{1-} \\ \psi_{1+} \\ \psi_{2-} \\ \psi_{2+} \end{pmatrix} \,,
 }
where the components $\psi_{\alpha\pm}$ depend only on $r$, and $g_{mn}$ is the part of the metric parallel to the boundary.  

Following the methods of \cite{Gubser:2012yb}, it is possible to reduce both \eno{DiracMostlyMinus} and \eno{DiracFour} to the form
 \eqn{XYZForm}{
  (\partial_r + X \sigma_3 + Y i \sigma_2 + Z \sigma_1) \psi_\alpha = 0 \,,
 }
where $\alpha=1,2$, and the functions $X$, $Y$, and $Z$ are real.  In five dimensions, we have
 \eqn{XYZexplicit}{
  X = {e^B m \over \sqrt{h}} \qquad\qquad
  Y = -{e^{-A+B} u \over \sqrt{h}} \qquad\qquad
  Z = -{e^{-A+B} \over \sqrt{h}} ((-1)^\alpha k - v)
 }
where
 \eqn[c]{uvmexplicit}{
  u = {\omega + g q_1 \Phi_1 + g q_2 \Phi_2 \over \sqrt{h}} \qquad\qquad
  v = 2 e^{-B} (p_1 e^{-2\phi / \sqrt{6}} \partial_r \Phi_1 +
     p_2 e^{\phi / \sqrt{6}} \partial_r \Phi_2)  \cr
  m = g (m_1 e^{-\phi/\sqrt{6}} + m_2 e^{2\phi/\sqrt{6}}) \,.
 }
In four dimensions, we have instead
 \eqn{XYZFour}{
  X = -{e^B \over 4 \sqrt{h}} \sum_i m_i e^{\lambda_i/2} \qquad\qquad
  Y = -{e^{-A+B} u \over \sqrt{h}} \qquad\qquad
  Z = -{e^{-A+B} \over \sqrt{h}} ((-1)^\alpha k - v)
 }
where
 \eqn[c]{uvmexplicitFour}{
  u = {\omega + {1 \over 4} \sum_i q_i \Phi_i \over \sqrt{h}} \qquad\qquad
  v = {e^{-B} \over 4} \sum_i p_i e^{-\lambda_i/2} \partial_r \Phi_i  \cr
 }

\subsection{Calculating the Green's function}

The standard approach to extracting a Green's function is to numerically solve \eno{XYZForm} subject to infalling boundary conditions near $r=r_H$ and then find $G_R$ as a ratio of coefficients characterizing the asymptotic behavior at large $r$.  Let's review this method before introducing the one based on fermion bilinears which we actually used for computation.

We will focus on two out of the four components of the spinor:
 \eqn{PsiComponents}{
  \psi_\alpha = \begin{pmatrix} \psi_{\alpha-} \\ \psi_{\alpha+} \end{pmatrix} \,,
 }
for a fixed value of $\alpha$.  For brevity, let us now suppress the index $\alpha$.  To leading order, the asymptotic behavior near the horizon is
 \eqn{PsiAsymptotic}{
  \psi_- = i \psi_+ = {i \over 2} (r-r_H)^{-{i\omega \over 4\pi T}} \,,
 }
and corrections arise at relative order $\sqrt{r-r_H}$.  The overall normalization of the wavefunction has been chosen for later convenience.  In the five-dimensional case, near the boundary of $AdS_5$, the asymptotics depends on the mass of the fermion, which for us is $m=1/2$.  In this case one finds
 \eqn{PsiFar}{
  \psi_+ &= A_\psi \sqrt{r} + {\cal O}(r^{-3/2} \log r)  \cr
  \psi_- &= A_\psi \left[ 
   (-1)^\alpha k + \omega + q_1 \mu_1 + {q_2 \mu_2 \over \sqrt{2}} \right] 
     {\log \mu r \over \sqrt{r}} + {D_\psi \over \sqrt{r}} + 
     {\cal O}(r^{-5/2} \log r) \,.
 }
Here $\mu_1$ and $\mu_2$ are the chemical potentials mentioned below \eno{BackgroundMetric}, while $\mu$ is an arbitrary energy scale and $A_\psi$ and $D_\psi$ are the asymptotic coefficients which we are after.  The Green's function is\footnote{We are discarding an overall factor from the Green's function which includes a factor of $1/G$ where $G$ is Newton's constant.  Restoring this factor would cause  $G_R$ to scale as $N^2$ in the $AdS_5$ case and $N^{3/2}$ in the $AdS_4$ case.}
 \eqn{GreenRatio}{
  G_R = {D_\psi \over A_\psi} \,.
 }
It is retarded because of the infalling boundary conditions.  The arbitrariness of $\mu$ translates into an arbitrariness in $G_R$ toward adding some real multiple of $(-1)^\alpha k + \omega + q_1 \mu_1 + {q_2 \mu_2 \over \sqrt{2}}$.  In real space this is a contact term.  The spectral weight $\Im G$ is unaffected by this ambiguity.

Recall that we suppressed the index $\alpha$.  When we restore it, the Green's function becomes a bispinor, $G^R_{\alpha\beta}$, which for our choice of Clifford basis is diagonal when $\vec{k}$ points in the $x^1$ direction.  Since the equation of motion \eno{XYZForm} depends on $k$ and the index $\alpha$ only through the combination $(-1)^\alpha k$, we may derive the relation $G^R_{22}(\omega,k) = G^R_{11}(\omega,-k)$.  For no special reason, we will focus on $G^R_{22}$ in our numerical computations.  $G^R_{22}$ is generally not a symmetric function of $k$; however, because of the relation just mentioned, Fermi surface behavior at $k=k_F$ in $G^R_{22}$ will be matched by Fermi surface behavior at $k=-k_F$ in $G^R_{11}$.  Indeed, spatial rotational symmetry is unbroken in the backgrounds we study, so any Fermi surfaces found must be spherical.

In the four-dimensional case, the near-horizon asymptotics are given as in \eno{PsiComponents}, but the near-boundary asymptotics are different.  The mass term vanishes at the boundary for the fermions we are interested in, so the appropriate asymptotics are
 \eqn{PsiFarFour}{
  \psi_+ = A_+ + {B_+ \over r} + {\cal O}(r^{-2}) \qquad\qquad
  \psi_- = A_- + {B_- \over r} + {\cal O}(r^{-2}) \,.
 }
The analysis of \cite{DeWolfe:2014ifa} shows that the correct Green's function of the supersymmetric M2-brane theory is 
 \eqn{GRRatioFour}{
  G_R = {A_- \over A_+}
 }
for all the cases we will study, as opposed to the expression $A_+ / A_-$ which follows from an alternative quantization.

In order to introduce the approach we actually use for computing $G_R$, we first define
 \eqn{Udefs}{
  U_+ = \psi_- + i \psi_+ \qquad\qquad
  U_- = \psi_- - i \psi_+ \,.
 }
Then \eno{XYZForm} immediately implies
 \eqn{FConserved}{
  \partial_r {\cal F} = 0 \qquad\hbox{where}\qquad
    {\cal F} = |U_+|^2 - |U_-|^2 \,.
 }
Slightly less obvious is the following set of definitions:
 \eqn{GIJDefs}{
  {\cal I} = U_+ U_-^* + U_+^* U_- \qquad\qquad
  {\cal J} = i(U_+ U_-^* - U_+^* U_-) \qquad\qquad
  {\cal K} = |U_+|^2 + |U_-|^2 \,.
 }
As a consequence of \eno{XYZForm}, these generalized fluxes obey
 \eqn{GIJeoms}{
  \partial_r \begin{pmatrix} {\cal I} \\ {\cal J} \\ {\cal K} \end{pmatrix} = 
    \begin{pmatrix} 0 & 2Y & -2X \\ -2Y & 0 & 2Z \\ -2X & 2Z & 0 \end{pmatrix}
    \begin{pmatrix} {\cal I} \\ {\cal J} \\ {\cal K} \end{pmatrix} \,.
 }
As a practical matter, it is more efficient to integrate the three real coupled equations \eno{GIJeoms} than the original spinorial equations \eno{XYZForm}.  This is partly because the asymptotic boundary conditions near the horizon corresponding to \eno{PsiAsymptotic} are non-oscillatory:
 \eqn{IJKNear}{
  {\cal I} = i_1 \sqrt{r-r_H} \qquad\qquad
  {\cal J} = j_1 \sqrt{r-r_H} \qquad\qquad
  {\cal K} = 1 \,,
 }
with additive corrections to all three quantities coming in at order $r-r_H$.  The coefficients $i_1$ and $j_1$ can be determined in terms of parameters of the differential equation, namely $r_H$, the charge parameters, $\omega$, $k$, and the parameters listed in tables~\ref{FiveDTable} and~\ref{FourDTable}.  Note also that \eno{PsiAsymptotic} corresponds to ${\cal F}=1$.  The non-oscillatory asymptotics \eno{IJKNear} contrasts with the strongly oscillatory behavior \eno{PsiAsymptotic}, which requires many steps in $r$ to accurately track.

The near-boundary asymptotics depends on the mass and the dimension.  In the five-dimensional case with $m=1/2$, one finds
 \eqn{IJKFar}{
  {\cal I} &= -K_{-1} r + {\cal O}(r^{-1} (\log r)^2)  \cr
  {\cal J} &= -2 K_{-1} \left[ (-1)^\alpha k + \omega + q_1 \mu_1 + 
    {q_2 \mu_2 \over \sqrt{2}} \right] \log \mu r + J_0 + 
    {\cal O}(r^{-2} (\log r)^2)  \cr
  {\cal K} &= K_{-1} r + {\cal O}(r^{-1} (\log r)^2) \,.
 }
Subleading terms at orders ${(\log r)^2 \over r}$ and ${\log r \over r}$ in ${\cal I}$ and ${\cal K}$, and at orders ${(\log r)^2 \over r^2}$ and ${\log r \over r^2}$ in ${\cal J}$, can be determined in terms of the integration constants $K_{-1}$ and $J_0$.  One more free integration constant comes up in the series solution: It is a contribution $K_1/r$ to ${\cal K}$, with similar pure power contributions to ${\cal I}$ and ${\cal J}$ determined in terms of $K_1$ as well as $K_{-1}$ and $J_0$.  By matching a series expansion of the form \eno{IJKFar} to a numerical solution to the current equations \eno{GIJeoms}, one can extract numerical values for $K_{-1}$, $J_0$, and $K_1$.  Note that $J_0$ suffers from the same sort of additive ambiguity that afflicts $D_\psi$, due to the arbitrariness of the scale $\mu$ occurring in the logarithm shown explicitly in \eno{IJKFar}.  This is not surprising given that we can use the far region expansions and \eno{PsiFar} and \eno{IJKFar} to relate
 \eqn{IJKcoefs}{
  {\cal F} = 2i (D_\psi^* A_\psi - D_\psi A_\psi^*) = 1 \qquad\qquad
  K_{-1} = 2 |A_\psi|^2 \qquad\qquad
  J_0 = -2 (A_\psi^* D_\psi + A_\psi D_\psi^*) \,.
 }
The last expression for ${\cal F}$ comes from using the near region expansion \eno{IJKFar}, which of course is allowed because ${\cal F}$ is exactly constant as a consequence of the equations of motion \eno{XYZForm}.  It follows immediately that
 \eqn{GRatios}{
  \Im G_R &= {1 \over 2i} {A_\psi^* D_\psi - A_\psi D_\psi^* \over |A_\psi|^2} = 
    {{\cal F} \over 2K_{-1}} = {1 \over 2 K_{-1}}  \cr
  \Re G_R &= {1 \over 2} {A_\psi^* D_\psi + A_\psi D_\psi^* \over |A_\psi|^2} = 
    -{J_0 \over K_{-1}} \,,
 }
where the last equality in the first line follows again from the flux at the horizon that we imposed through the choice \eno{PsiAsymptotic}.

In the four-dimensional case with $m=0$, the only thing that changes is the near-boundary asymptotics, which are simpler than before:
 \eqn{IJKFarFour}{
  {\cal I} = I_0 + {\cal O}(r^{-1}) \qquad
  {\cal J} = J_0 + {\cal O}(r^{-1}) \qquad
  {\cal K} = K_0 + {\cal O}(r^{-1}) \,.
 }
Expressing the fluxes in terms of the fermions, one finds
 \eqn{CoefFormsFour}{\seqalign{\span\TL & \span\TR &\qquad\qquad \span\TL & \span\TR}{
  {\cal F} &= 2i (A_-^* A_+ - A_+^* A_-) &
  I_0 &= 2 (|A_-|^2 - |A_+|^2)  \cr
  J_0 &= -2 (A_+^* A_- + A_-^* A_+) &
  K_0 &= 2 (|A_-|^2 + |A_+|^2) \,.
 }}
Recalling that $G_R = A_- / A_+$ for the cases we will study, we find
 \eqn{GreenFormFour}{
  \Im G_R &= {1 \over 2i} {A_+^* A_- - A_-^* A_+ \over |A_+|^2} = 
    {{\cal F} \over K_0 - I_0} =
    {1 \over K_0 - I_0}  \cr
  \Re G_R &= {1 \over 2} {A_+^* A_- + A_-^* A_+ \over |A_+|^2} = 
    -{J_0 \over K_0 - I_0} \,.
 }

\section{Examples and interpretation}
\label{NUMERICS}

\begin{table}
\centerline{
\begin{tabular}{c|c|c|c|c|c|c|c|c|c|c|} \hline
\# & $\chi^{q_a q_b q_c}$ &Dual operator &$m_1$& $m_2$ & $q_1$& $q_2$  & $p_1$ & $p_2$ & 1Q-5d & 2Q-5d \\ \hline
1 & $\chi^{({3 \over 2}, {1 \over 2}, {1 \over 2})}$ &$\lambda_1 Z_1$ &$-{1 \over 2}$& ${3 \over 4}$ & ${3 \over 2}$& $1$ & $- {1 \over 4}$& ${1 \over 2}$ & Y${}^{\rm 1A}$ & N${}^{\rm 1D}$ \\
2 & $\chi^{({3 \over 2}, -{1 \over 2}, -{1 \over 2})}$ & $\lambda_2 Z_1$&$-{1 \over 2}$& ${3 \over 4}$  & ${3 \over 2}$& $-1$ & $- {1 \over 4} $& $-{1 \over 2}$ & Y${}^{\rm 1A}$ & N${}^{\rm 1E}$ \\
3 & $\bar\chi^{({3 \over 2}, -{1 \over 2}, {1 \over 2})}$  , $\bar\chi^{({3 \over 2}, {1 \over 2}, -{1 \over 2})}$ & $\overline\lambda_3 Z_1$, $\overline\lambda_4 Z_1$&$-{1 \over 2}$& ${3 \over 4}$  & ${3 \over 2}$& $0$ & $-{1 \over 4} $&$0$ & Y${}^{\rm 1A}$ & N${}^{\rm 1F}$ \\[1pt] \hline
4 & $\chi^{({1 \over 2}, {3 \over 2}, {1 \over 2})}$, $\chi^{({1 \over 2}, { 1\over 2}, {3 \over 2})}$ & $\lambda_1 Z_2$, $\lambda_1 Z_3$&${1 \over 2}$& $-{1 \over 4}$  & ${1 \over 2}$& $2$ & ${1 \over 4} $& $0$ & N${}^{\rm 1B}$ & Y${}^{\rm 1G}$ \\
5 & $\bar\chi^{(-{1 \over 2},  {3 \over 2}, {1 \over 2})}$, $\bar\chi^{(-{1 \over 2},  {1 \over 2}, {3 \over 2})}$ & $\overline\lambda_2 Z_2$, $\overline\lambda_2 Z_3$&${1 \over 2}$& $-{1 \over 4}$ & $-{1 \over 2}$& $2$ & $-{1 \over 4}$& $0$ & N${}^{\rm 1C}$ & Y${}^{\rm 1G}$ \\ 
6 & $\chi^{(-{1 \over 2},  {3 \over 2}, -{1 \over 2})}$, $\chi^{(-{1 \over 2}, -{1 \over 2},  {3 \over 2})}$& $\lambda_3 Z_2$, $\lambda_4 Z_3$&${1 \over 2}$& $-{1 \over 4}$& $-{1 \over 2}$& $1$ & $- {1 \over 4}$& $- {1 \over 2}$ & N${}^{\rm 1C}$ & Y${}^{\rm 1H}$ \\
7 & $\bar\chi^{({1 \over 2}, -{1 \over 2},  {3 \over 2})}$, $\bar\chi^{({1 \over 2},  {3 \over 2}, -{1 \over 2})}$& $\overline\lambda_3 Z_3$, $\overline\lambda_4 Z_2$& ${1 \over 2}$& $-{1 \over 4}$ & ${1 \over 2}$& $1$ & ${1 \over 4}$& $-{1 \over 2}$ & N${}^{\rm 1B}$ & Y${}^{\rm 1H}$ \\[1pt] \hline
\end{tabular}
}
 \caption{Presence or absence of holographic Fermi surfaces in five-dimensional maximal gauged supergravity in the one-charge and two-charge backgrounds.  All the cases listed have $m=1/2$, so the analysis leading to \eno{GRatios} applies.  The superscripts in the last two columns indicate the figure in which the relevant spectral weight is plotted.}\label{FiveDTable}
\end{table}

\def\s#1{${}^{\rm 2#1}$}
\def\t#1{${}^{\rm 3#1}$}
\begin{table}
\centerline{
\begin{tabular}{c|c||c|c|c|c||c|c|c|c||c|c|c|} \hline
\# & Active boson & $q_a$ & $q_b$ & $q_c$ & $q_d$ & $m_a$ & $m_b$ & $m_c$ & $m_d$ &
 1Q-4d & 2Q-4d & 3Q-4d  \\ \hline
1 & $Z_1$ & $3$ & $-1$ & $1$ & $1$ & $-3$ & $1$ & $1$ & $1$ & Y\s{A} & N\s{D} & N\t{I}  \\
2 & $Z_1$ & $3$ & $1$ & $-1$ & $1$ & $-3$ & $1$ & $1$ & $1$ & Y\s{A} & N\s{E} & N\t{I}  \\
3 & $Z_1$ & $3$ & $1$ & $1$ & $-1$ & $-3$ & $1$ & $1$ & $1$ & Y\s{A} & N\s{E} & N\t{I}  \\
4 & $Z_2$ & $-1$ & $3$ & $1$ & $1$ & $1$ & $-3$ & $1$ & $1$ & N\s{B} & N\s{D} & Y\t{J}  \\
5 & $Z_2$ & $1$ & $3$ & $-1$ & $1$ & $1$ & $-3$ & $1$ & $1$ & N\s{C} & N\s{E} & Y\t{K}  \\
6 & $Z_2$ & $1$ & $3$ & $1$ & $-1$ & $1$ & $-3$ & $1$ & $1$ & N\s{C} & N\s{E} & Y\t{K}  \\
7 & $Z_3$ & $-1$ & $1$ & $3$ & $1$ & $1$ & $1$ & $-3$ & $1$ & N\s{B} & Y\s{F} & Y\t{J}  \\
8 & $Z_3$ & $1$ & $-1$ & $3$ & $1$ & $1$ & $1$ & $-3$ & $1$ & N\s{C} & Y\s{F} & Y\t{K}  \\
9 & $Z_3$ & $1$ & $1$ & $3$ & $-1$ & $1$ & $1$ & $-3$ & $1$ & N\s{C} & Y\s{G} & Y\t{K}  \\
10 & $Z_4$ & $-1$ & $1$ & $1$ & $3$ & $1$ & $1$ & $1$ & $-3$ & N\s{B} & Y\s{F} & Y\t{J}  \\
11 & $Z_4$ & $1$ & $-1$ & $1$ & $3$ & $1$ & $1$ & $1$ & $-3$ & N\s{C} & Y\s{F} & Y\t{K}  \\
12 & $Z_4$ & $1$ & $1$ & $-1$ & $3$ & $1$ & $1$ & $1$ & $-3$ & N\s{C} & Y\s{G} & Y\t{K}  \\ \hline
13 & $Z_1$ & $3$ & $-1$ & $-1$ & $-1$ & $-3$ & $1$ & $1$ & $1$ & Y\s{A} & N\s{H} & N\t{L}  \\
14 & $Z_2$ & $-1$ & $3$ & $-1$ & $-1$ & $1$ & $-3$ & $1$ & $1$ & N\s{B} & N\s{H} & Y\t{M}  \\
15 & $Z_3$ & $-1$ & $-1$ & $3$ & $-1$ & $1$ & $1$ & $-3$ & $1$ & N\s{B} & Y\s{G} & Y\t{M}  \\
16 & $Z_4$ & $-1$ & $-1$ & $-1$ & $3$ & $1$ & $1$ & $1$ & $-3$ & N\s{B} & Y\s{G} & Y\t{M}  \\ \hline
\end{tabular}
}
\caption{Presence or absence of holographic Fermi surfaces in four-dimensional maximal gauged supergravity in the one-, two-, and three-charge backgrounds.  The values of the four Pauli couplings are $p_i = m_i/q_i$ in every case.  All the cases listed have vanishing total mass near the boundary, and the analysis leading to \eno{GreenFormFour} applies.  The superscripts in the last three columns indicate the figure in which the relevant spectral weight is plotted.}\label{FourDTable}
\end{table}

Armed with the fermion bilinear methodology explained in \eno{Udefs}-\eno{GIJeoms} and the formulas \eno{GRatios} for five-dimensional cases and \eno{GreenFormFour} for four-dimensional cases, we computed finite-temperature two-point Green's functions for the operators dual to the supergravity fermions listed in table~\ref{FiveDTable} for five-dimensional maximal gauged supergravity, and in table~\ref{FourDTable} for four-dimensional maximal gauged supergravity.

In table~\ref{FiveDTable}, of which all but the last two columns are taken directly from \cite{DeWolfe:2012uv}, we have identified the $U(1)^3$ charges of each fermion, $(q_a,q_b,q_c)$, such that $q_1 = q_a$ and $q_2 = q_b+q_c$.  The dual operators in ${\cal N}=4$ super-Yang-Mills are also identified in table~\ref{FiveDTable}, using a standard complex notation for the adjoint scalars, namely $Z_1 = X_1 + i X_2$, $Z_2 = X_3 + i X_4$, and $Z_3 = X_5 + i X_6$.  These complex combinations have weights $(1,0,0)$, $(0,1,0)$, and $(0,0,1)$ under $SO(6)$.  In a Weyl basis, the gauginos $\lambda$ carry quantum numbers in the ${\bf 4}$ of $SO(6)$, namely $\left( {1 \over 2}, {1 \over 2}, {1 \over 2} \right)$ for $\lambda_1$, $\left( {1 \over 2}, -{1 \over 2}, -{1 \over 2} \right)$ for $\lambda_2$, $\left( -{1 \over 2}, {1 \over 2}, -{1 \over 2} \right)$ for $\lambda_3$, and $\left( -{1 \over 2}, -{1 \over 2}, {1 \over 2} \right)$ for $\lambda_4$.  The point to note is that one complex scalar or another is ``active'' in each case, meaning that it is included in the fermionic operator.

In table~\ref{FourDTable}, all but the last three columns are from \cite{DeWolfe:2014ifa}, and we have identified in each case which boson is ``active'' in the same sense as before, using the complex notation $Z_j = X_{2j-1} + i X_{2j}$.  The fermionic superpartners of the $Z_j$ have $SO(8)$ weights of the form $\left( {3 \over 2},\pm {1 \over 2},\pm {1 \over 2},\pm {1 \over 2} \right)$ and permutations, with an odd number of minus signs.  We follow \cite{DeWolfe:2014ifa} in rescaling these weight vectors to $(q_a,q_b,q_c,q_d) = (3,\pm 1,\pm 1,\pm 1)$ and permutations, again with an odd number of minus signs.  The operator dual to any of the fermions in table~\ref{FourDTable} takes the form $\tr \lambda Z$ where the particular scalar one needs is determined by the position of $3$ in the rescaled weight vector---so $Z_1$ if $q_a = 3$, $Z_2$ if $q_b = 3$, and so on.

We examined two contrasting backgrounds in five dimensions and three in four dimensions.  A summary of these backgrounds is as follows:
 \begin{itemize}
  \item 1Q-5d refers to the planar one-charge black hole in $AdS_5$, with $Q_1=1$ and $Q_2=0$.  Inspection of the thermodynamics shows that as $r_H \to 0$, $T$ approaches a positive constant while $\mu_1 \to 0$, with entropy density $s \propto \mu_1^2$ and charge density $\rho_1 \propto \mu_1$.
  \item 2Q-5d refers to the planar two-charge black hole in $AdS_5$, with $Q_1=0$ and $Q_2=0$.  As $r_H \to 0$, $T \to 0$ and $\mu_2$ remains finite, while $s \propto T$ and $\rho_2$ remains finite.
  \item 1Q-4d refers to the planar one-charge black hole in $AdS_4$, with $Q_a=1$ and $Q_b=Q_c=Q_d=0$.  As $r_H \to 0$, $T \to 0$ and $\mu_a \to 0$ with $\mu_a \propto T^2$, $s \propto T^3$, and $\rho_a \propto T^2$.
  \item 2Q-4d refers to the planar two-charge black hole in $AdS_4$, with $Q_a=Q_b=0$ and $Q_c=Q_d=1$.  As $r_H \to 0$, $T$ remains finite and $\mu_d \to 0$, with $s \propto \mu_d^2$ and $\rho_d \propto \mu_d$, similar to the 1Q-5d background.
  \item 3Q-4d refers to the planar three-charge black hole in $AdS_4$, with $Q_a=0$ and $Q_b=Q_c=Q_d=1$.  As $r_H \to 0$, $T \to 0$ and $\mu_d$ remains finite, while $s \propto T$ and $\rho_d$ remains finite, similar to the 2Q-5d background.
 \end{itemize}
Note that all of these backgrounds have zero-entropy extremal limits; however, these limits are singular, at least in their description in maximal gauged supergravity.  The singularities involve divergences of the scalars as well as of curvature invariants, and one might worry that Green's function calculations directly in the extremal geometries would require some understanding beyond supergravity of how one should treat the boundary conditions of the fermion wave-function near the singularity.  By turning on small but non-zero $r_H$, we avoid this issue: the singularity is cloaked by a regular horizon, and the calculation is in a regime where supergravity is uniformly reliable, provided we think of taking the Planck length to $0$ (meaning $N$ to infinity) first, before sending $r_H \to 0$.

Because the underlying theories are conformal, one can only speak meaningfully of finite $\omega$ and $k$ as compared to some definite scale introduced by the background under consideration.  A convenient uniform choice in five dimensions is
 \eqn{MuFiveDef}{
  \mu_* = \sqrt{T^2 + \mu_1^2 + \mu_2^2} \,,
 }
while in four dimensions we used
 \eqn{MuFourDef}{
  \mu_* = \sqrt{T^2 + \mu_a^2 + \mu_b^2 + \mu_c^2 + \mu_d^2} \,.
 }
Because we always work at finite temperature, we can never expect a true Fermi surface singularity in the Green's function.  Instead, if there is a holographic Fermi surface, it should show up as a peak in the spectral weight $\Im G_R$ at $\omega=0$ whose width is no greater than some ${\cal O}(1)$ factor times the temperature.  This width criterion is important because essentially all the Green's functions we studied have a peak at some $k_*$, but some of the peaks are very broad compared to $T$, and we do not regard these as evidence of Fermi surfaces.  A confirmatory criterion, which we will term the $\omega$ criterion, is that if a maximum at $k=k_*$ is to be regarded as a Fermi surface, then the spectral weight as a function of $\omega$ for $k$ close to $k_*$ should show a peak with maximum close to $0$, consistent with a quasi-particle near the Fermi surface---though it may be hard to argue in some cases that this quasi-particle is long-lived.  A more heuristic criterion, which we will refer to as the magnitude criterion, is that the magnitude of the spectral weight should be large at $\omega=0$ and $k=k_*$ if it is to be regarded as a Fermi surface; more precise would be to say that if the maximum spectral weight at $\omega=0$ over any value of $k$ is parametrically suppressed, we probably do not have a Fermi surface.

\begin{figure}
\def\PaneWidth{3in}
\def\RowSpacing{0in}
\def\ColumnSpacing{0.5in}
\hbox{\includegraphics[width=\PaneWidth]{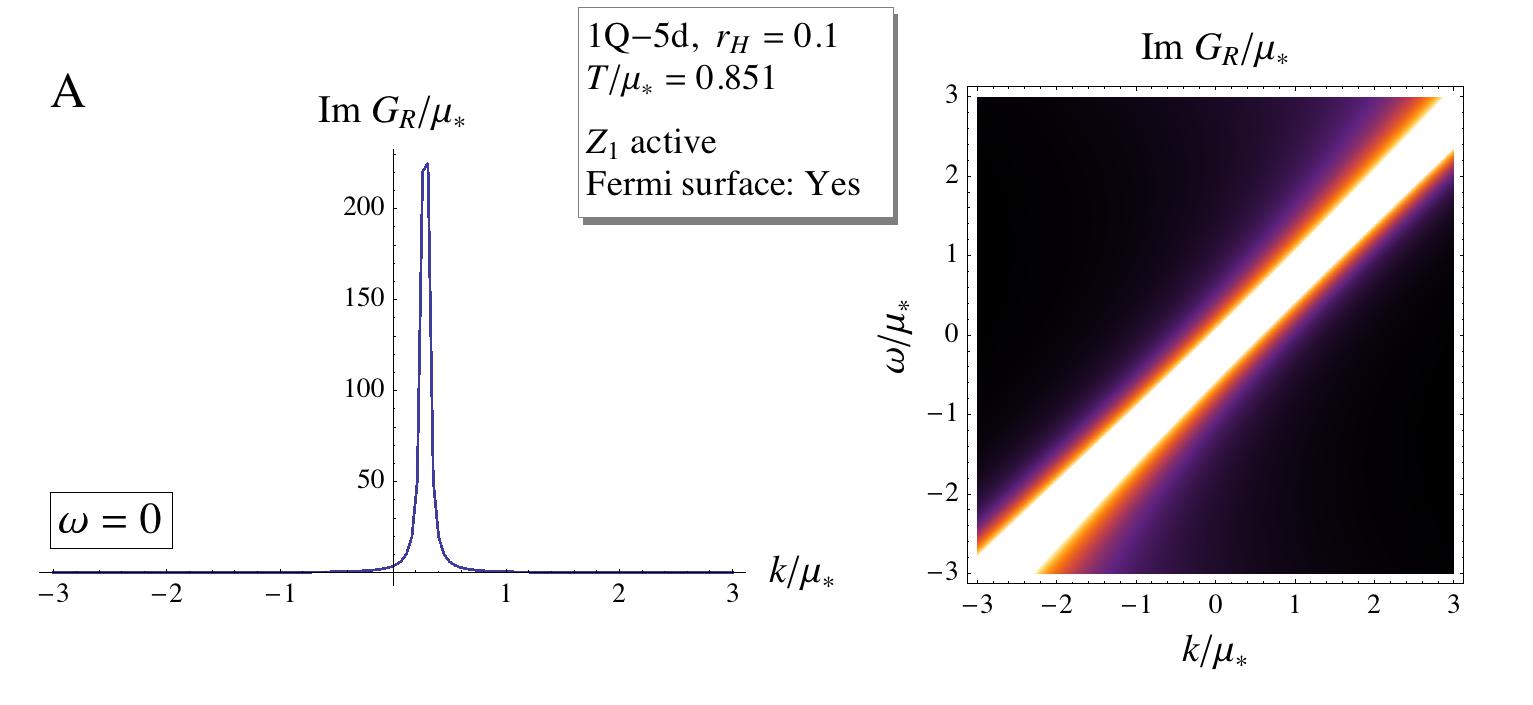}\hskip\ColumnSpacing\includegraphics[width=\PaneWidth]{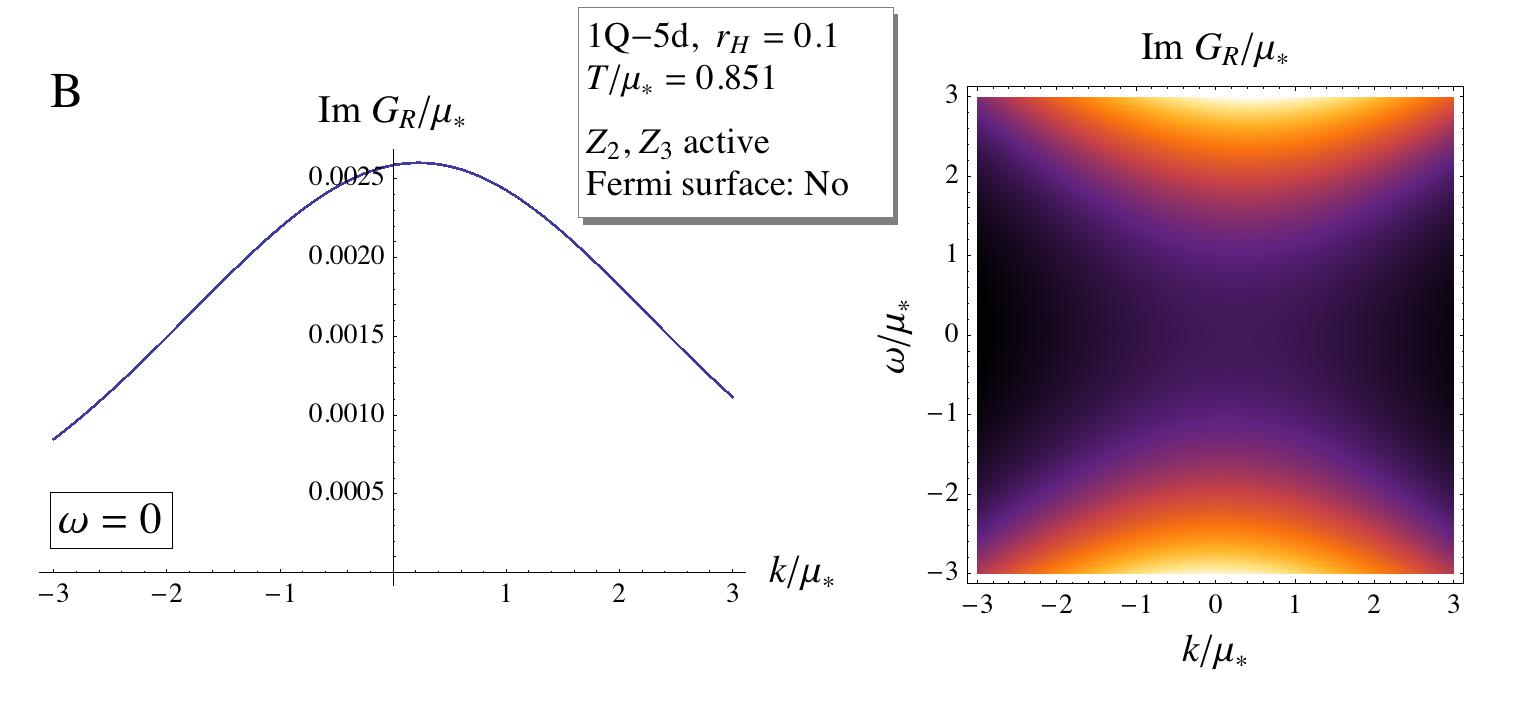}}\vskip\RowSpacing
\hbox{\includegraphics[width=\PaneWidth]{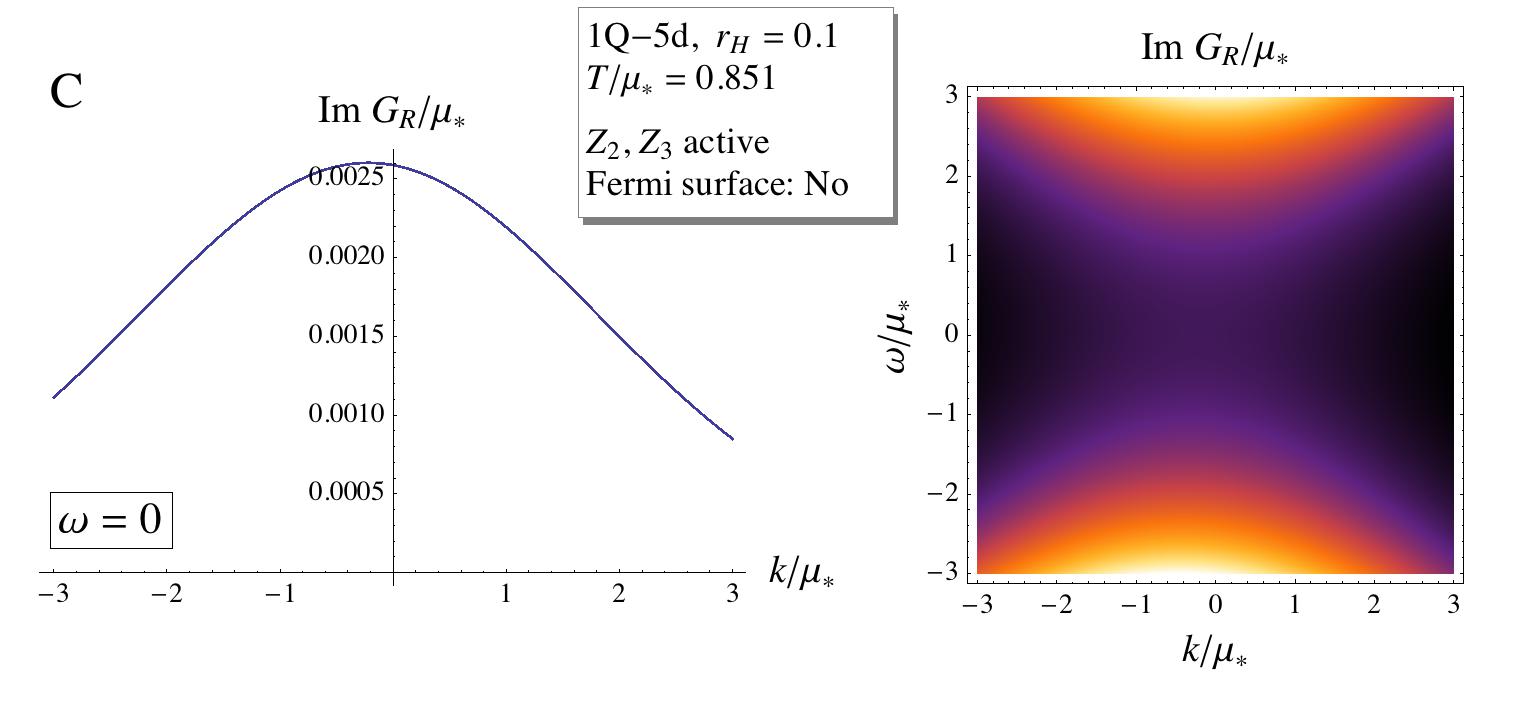}\hskip\ColumnSpacing\includegraphics[width=\PaneWidth]{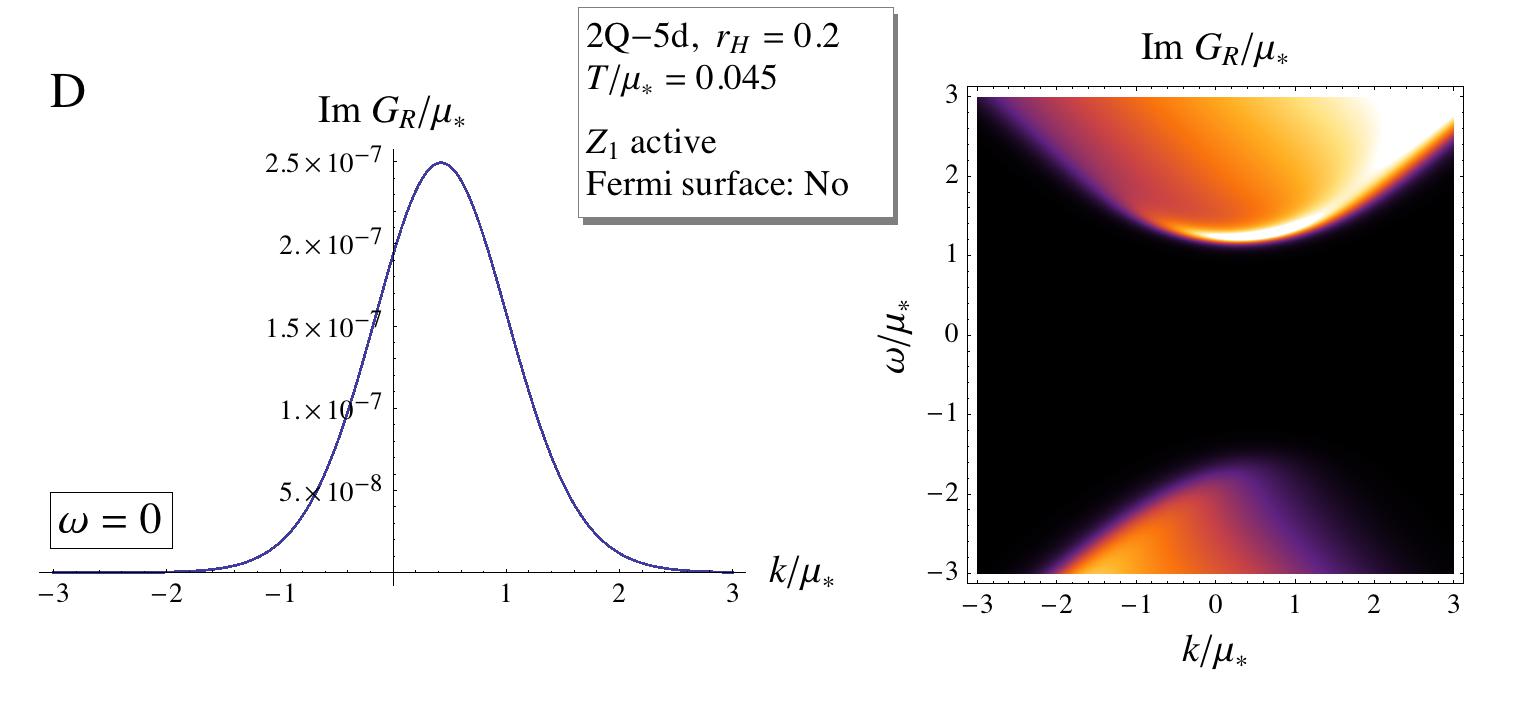}}\vskip\RowSpacing
\hbox{\includegraphics[width=\PaneWidth]{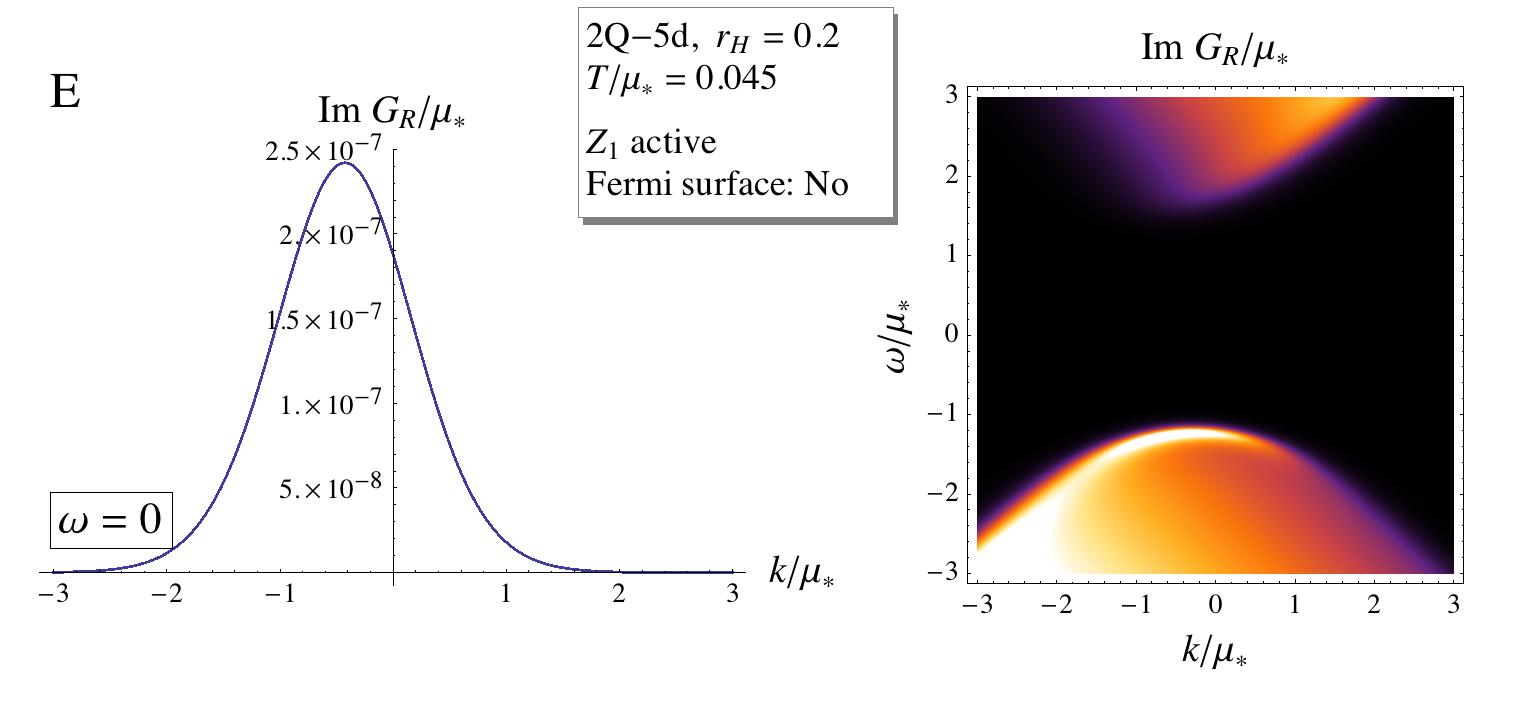}\hskip\ColumnSpacing\includegraphics[width=\PaneWidth]{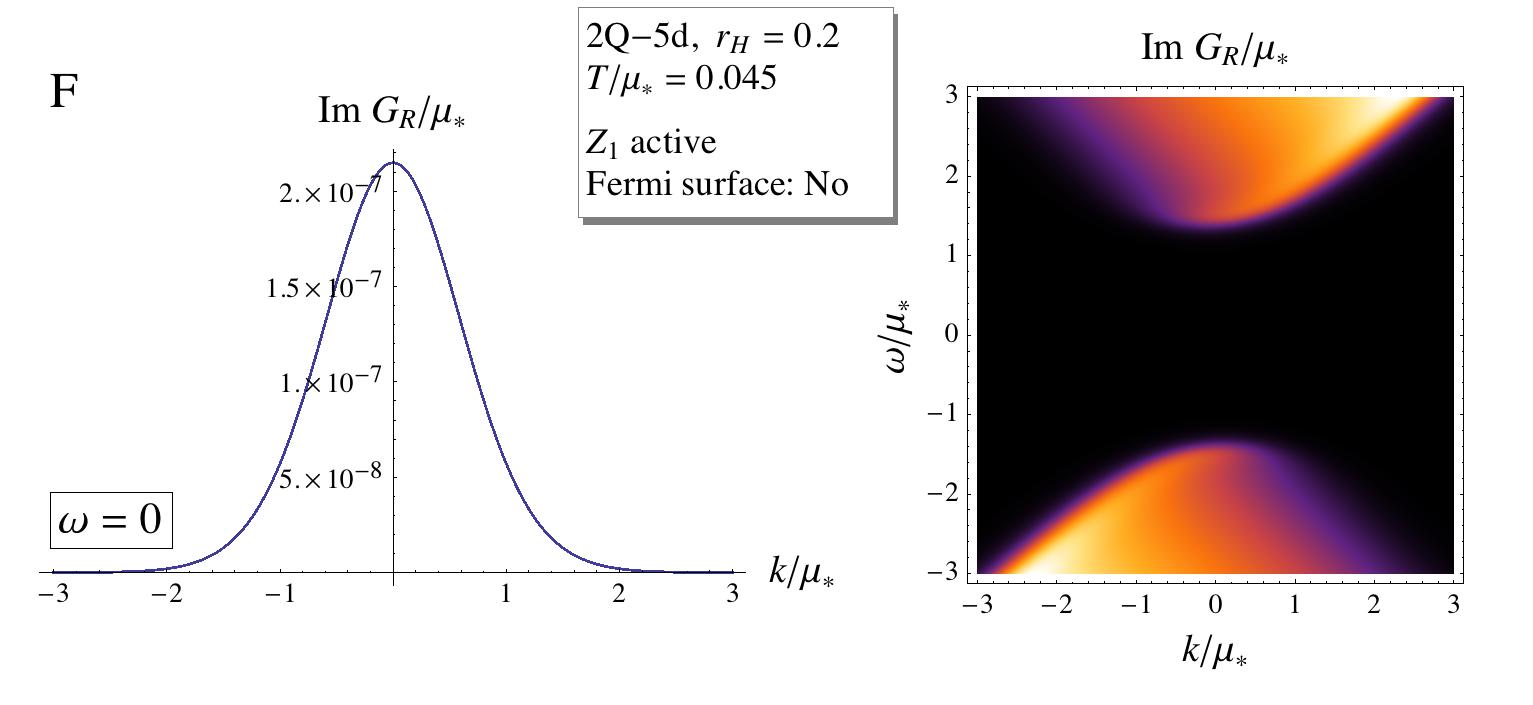}}\vskip\RowSpacing
\hbox{\includegraphics[width=\PaneWidth]{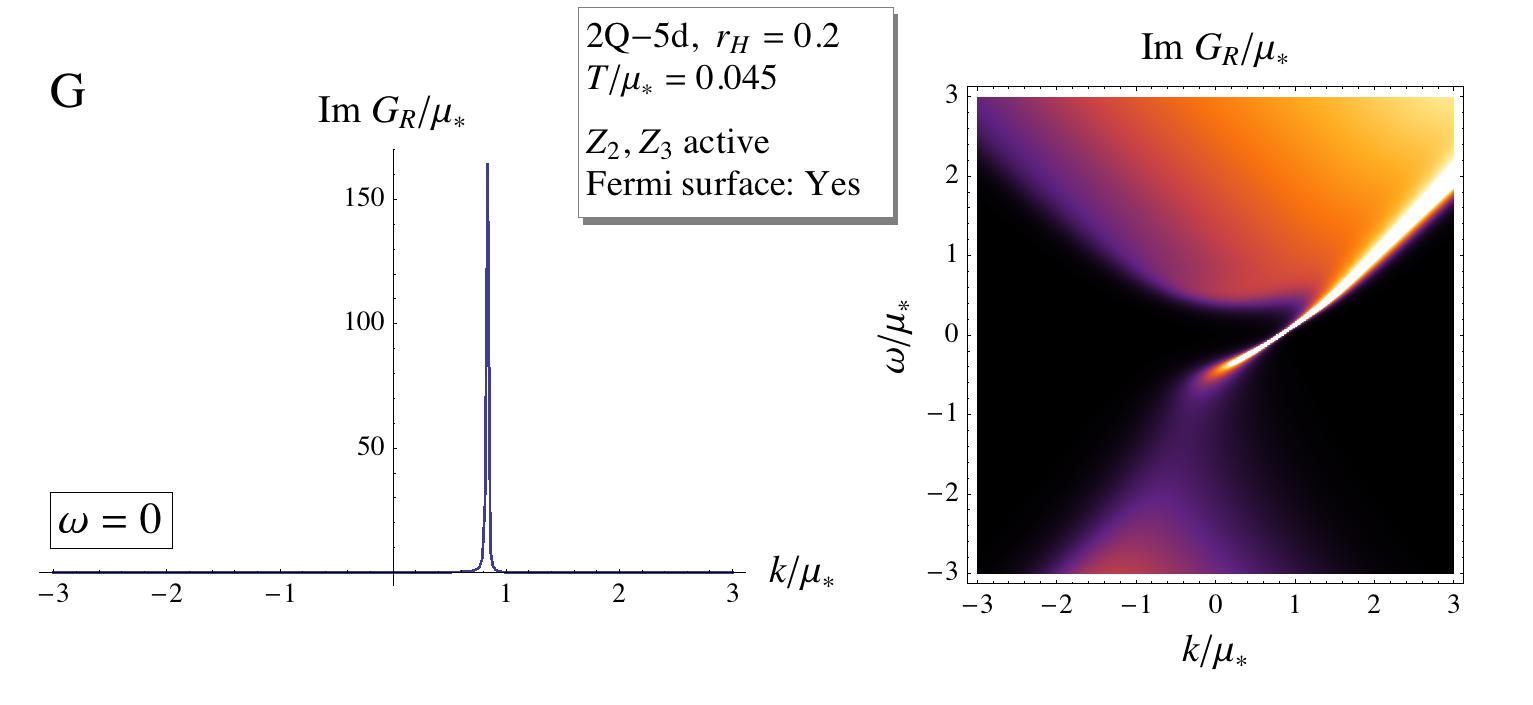}\hskip\ColumnSpacing\includegraphics[width=\PaneWidth]{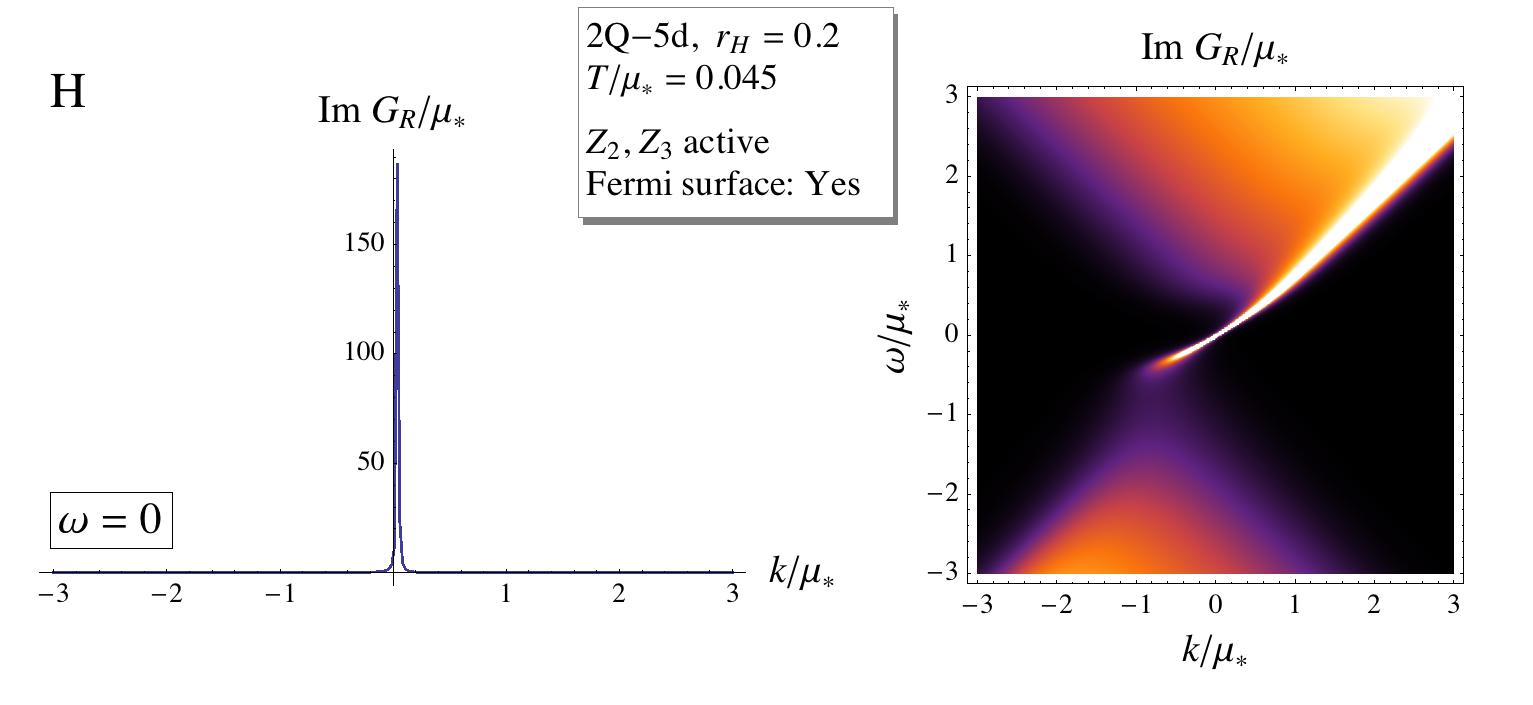}}\vskip\RowSpacing
\caption{The spectral weight of fermionic Green's functions in charged black hole backgrounds of five-dimensional gauged supergravity.  Accuracy of the height of the peaks in cases G and H is limited by the resolution on the $k$ axis.}\label{FiveDExamples}
\end{figure}

\begin{figure}
\def\PaneWidth{3in}
\def\RowSpacing{0in}
\def\ColumnSpacing{0.5in}
\hbox{\includegraphics[width=\PaneWidth]{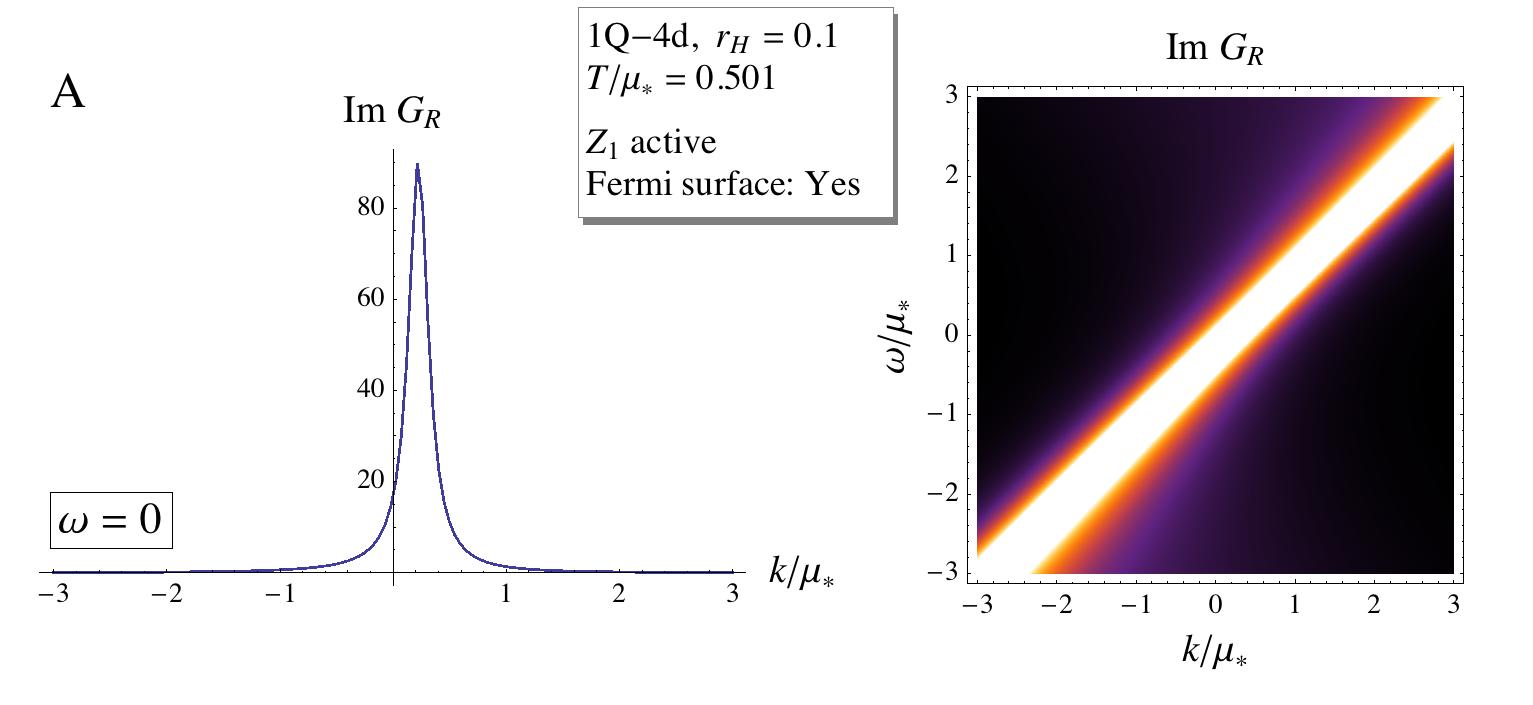}\hskip\ColumnSpacing\includegraphics[width=\PaneWidth]{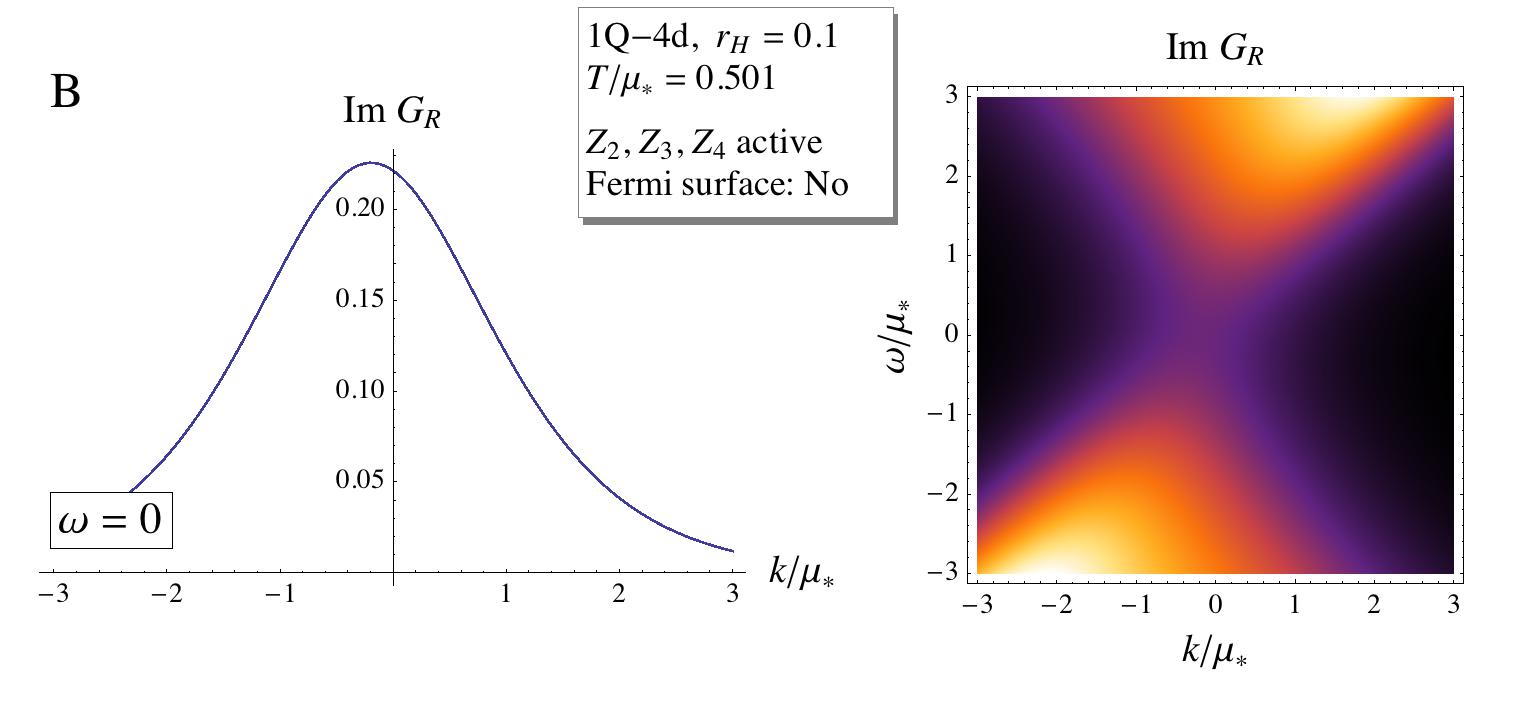}}\vskip\RowSpacing
\hbox{\includegraphics[width=\PaneWidth]{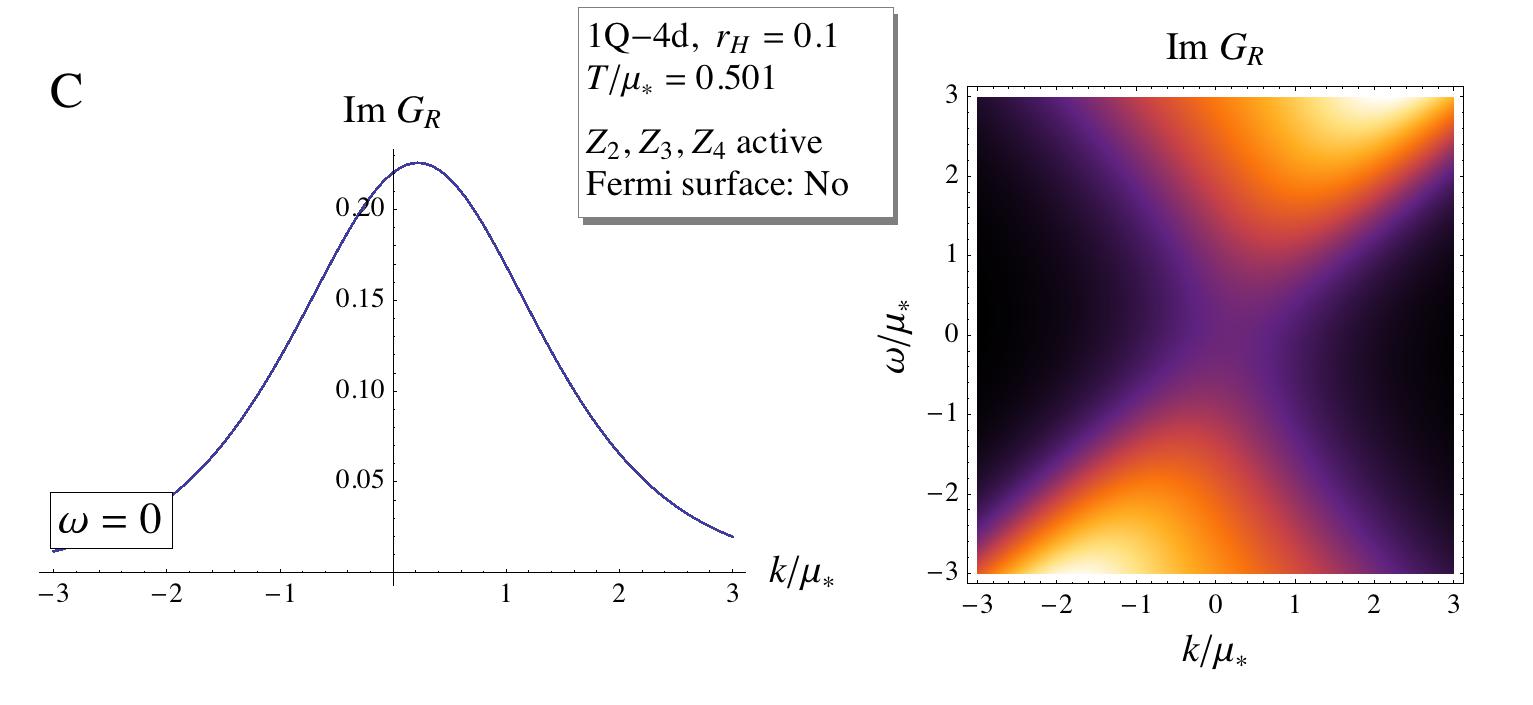}\hskip\ColumnSpacing\includegraphics[width=\PaneWidth]{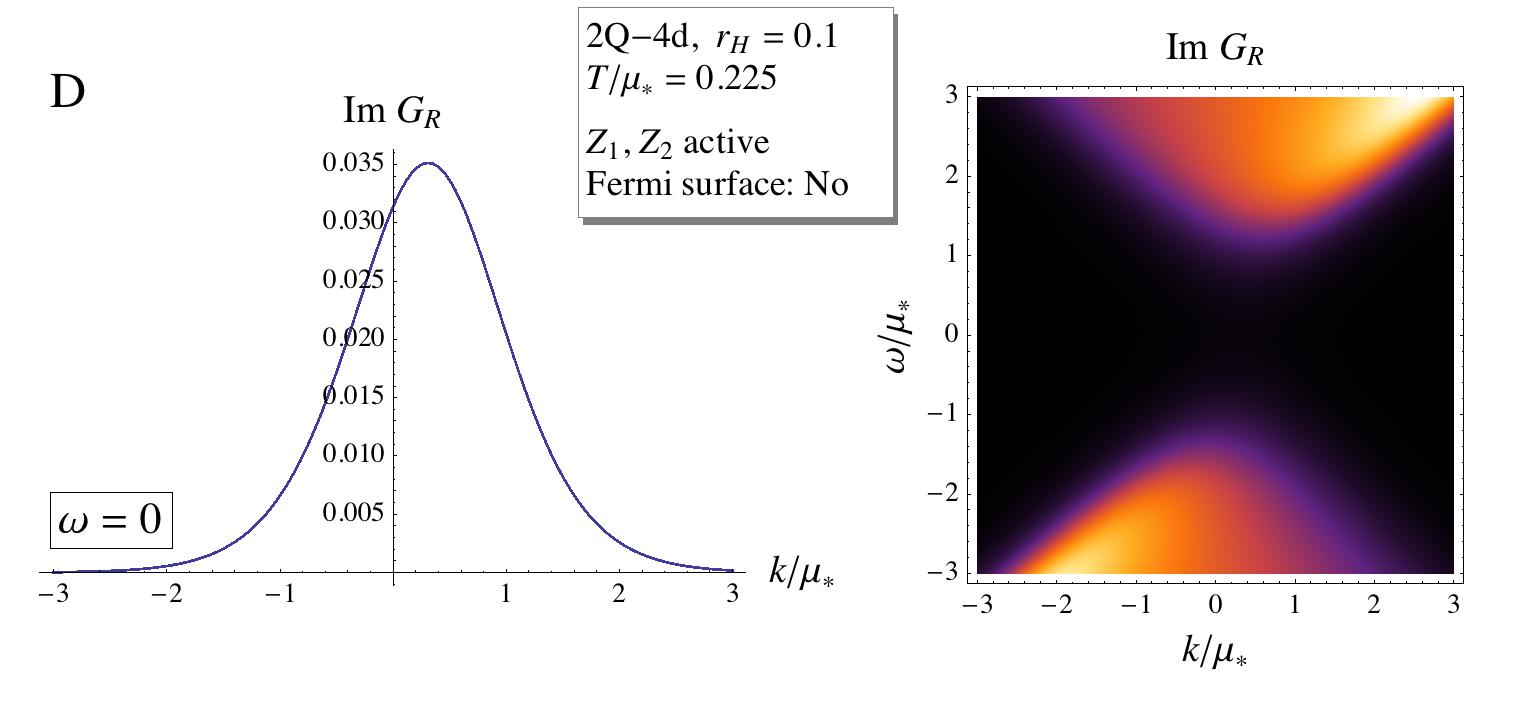}}\vskip\RowSpacing
\hbox{\includegraphics[width=\PaneWidth]{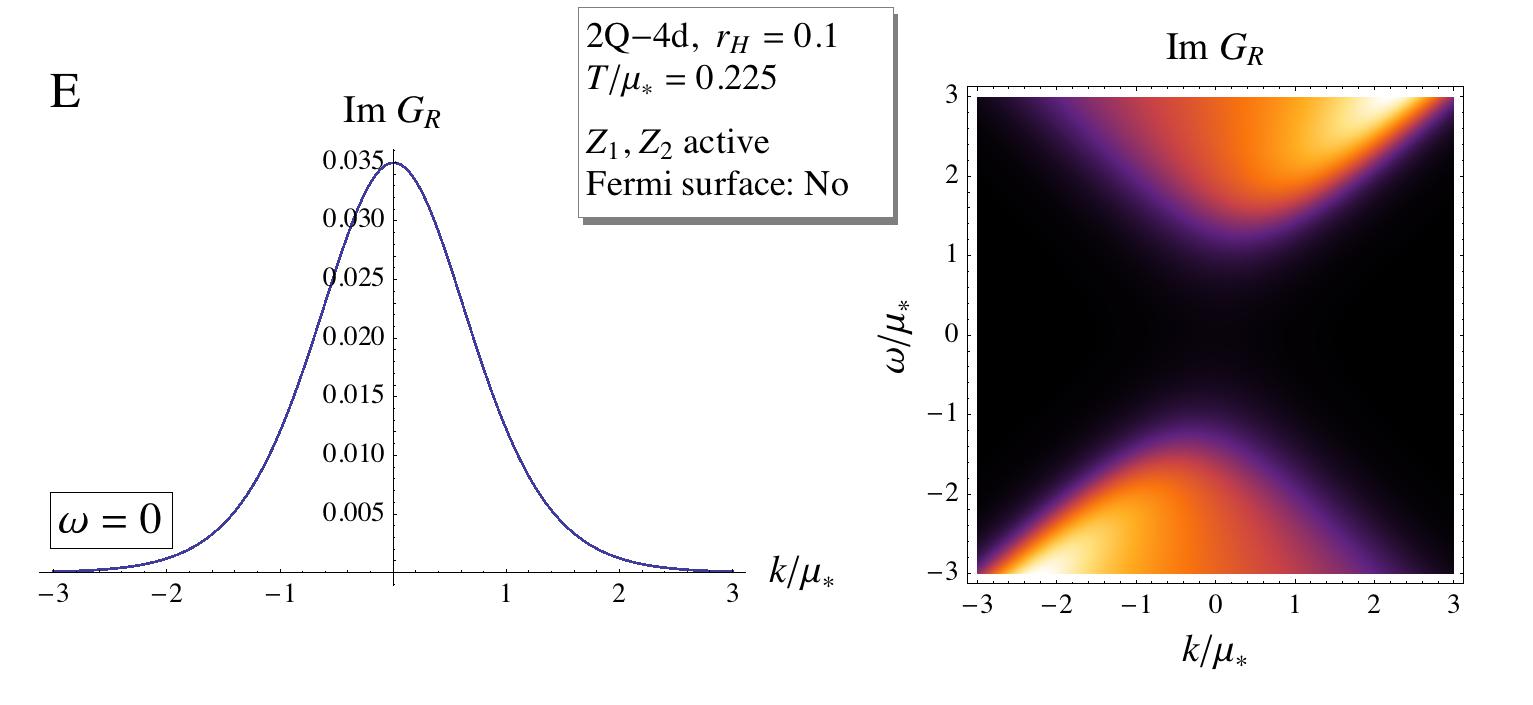}\hskip\ColumnSpacing\includegraphics[width=\PaneWidth]{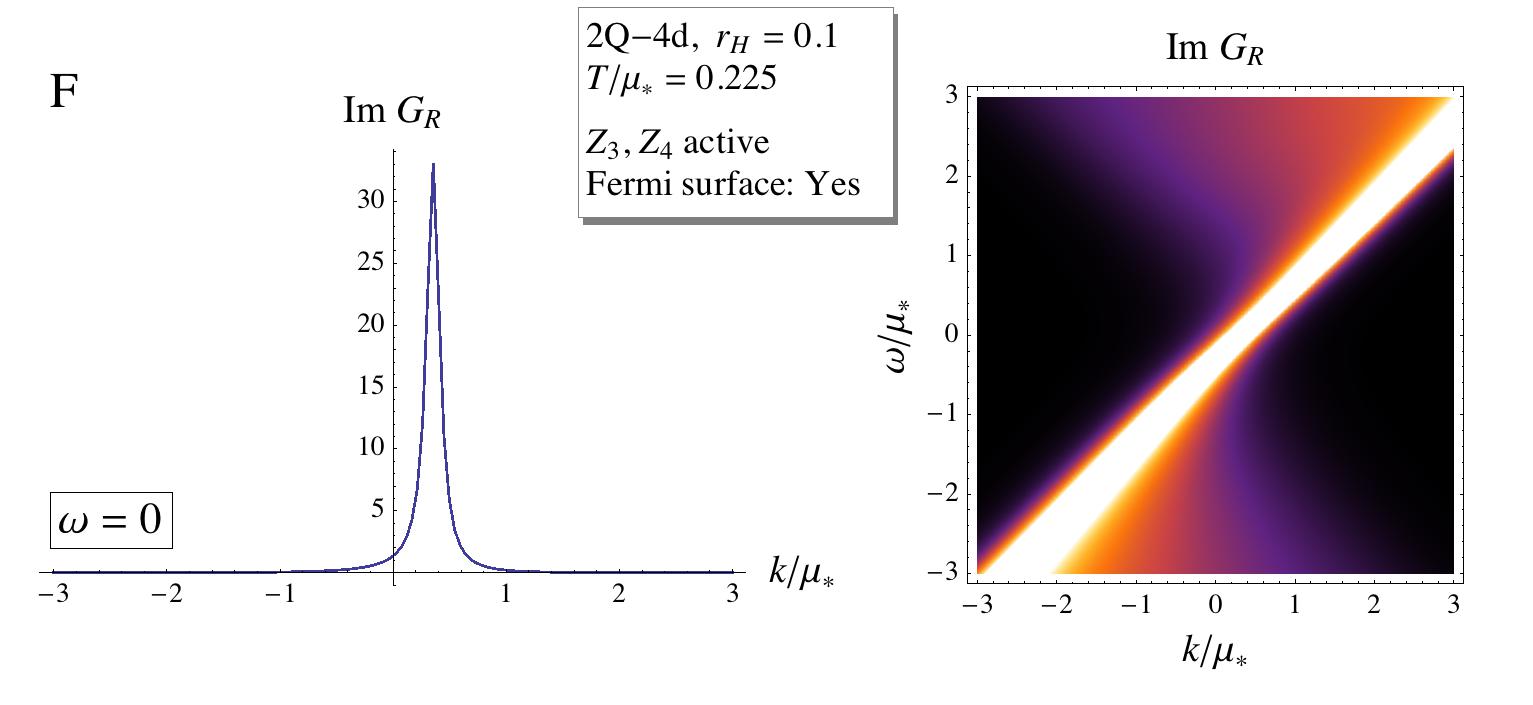}}\vskip\RowSpacing
\hbox{\includegraphics[width=\PaneWidth]{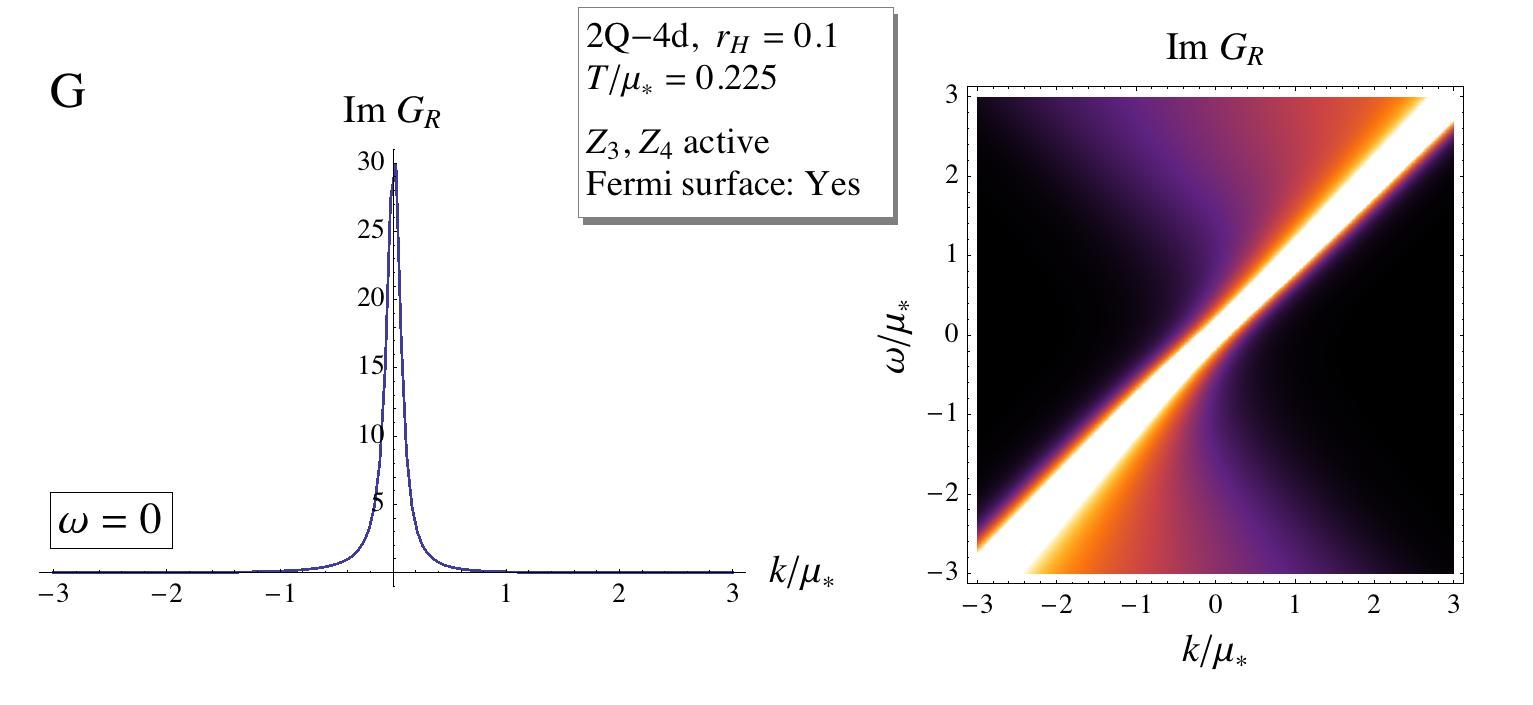}\hskip\ColumnSpacing\includegraphics[width=\PaneWidth]{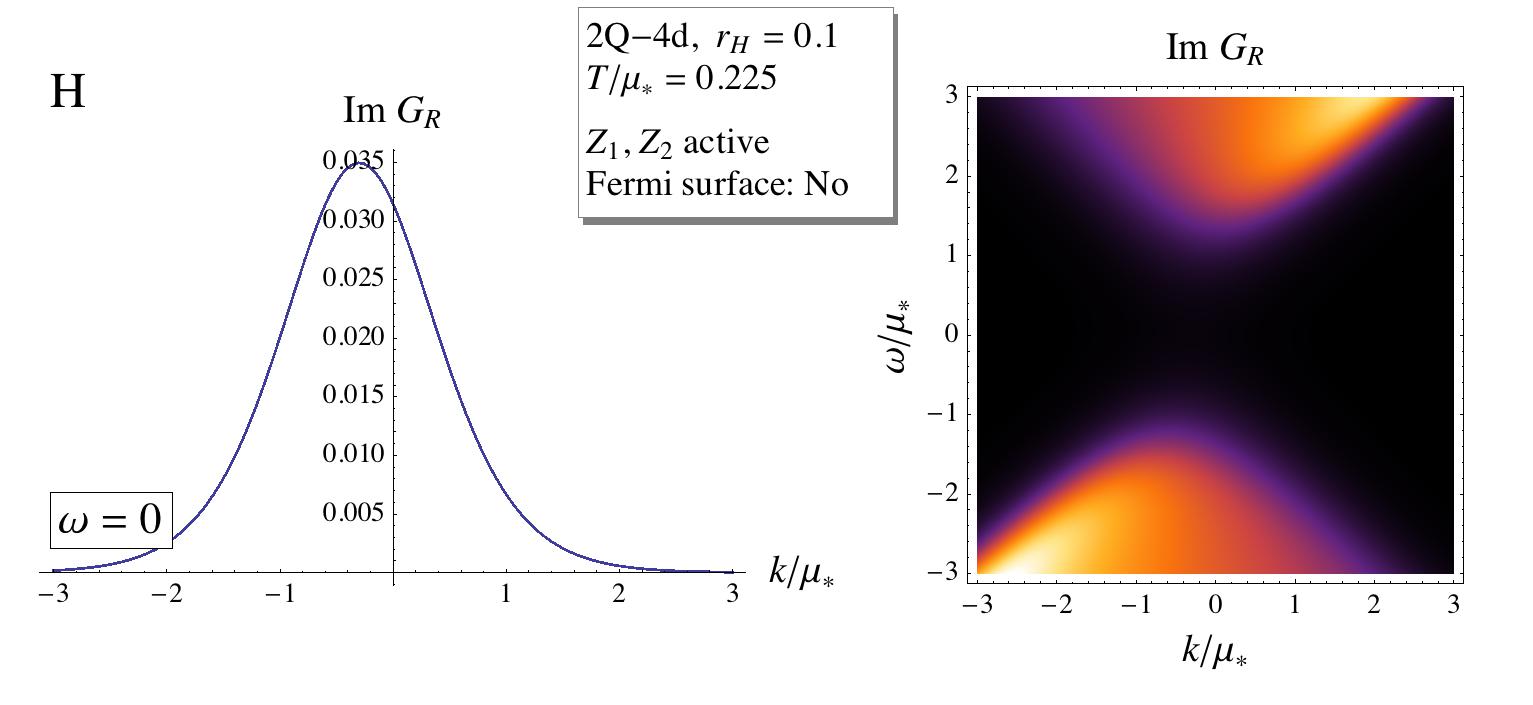}}\vskip\RowSpacing
\caption{The spectral weight of fermionic Green's functions in charged black hole backgrounds of five-dimensional gauged supergravity; continued in figure~\ref{FourDExamplesMore}.}\label{FourDExamples}
\end{figure}

\begin{figure}
\def\PaneWidth{3in}
\def\RowSpacing{0in}
\def\ColumnSpacing{0.5in}
\hbox{\includegraphics[width=\PaneWidth]{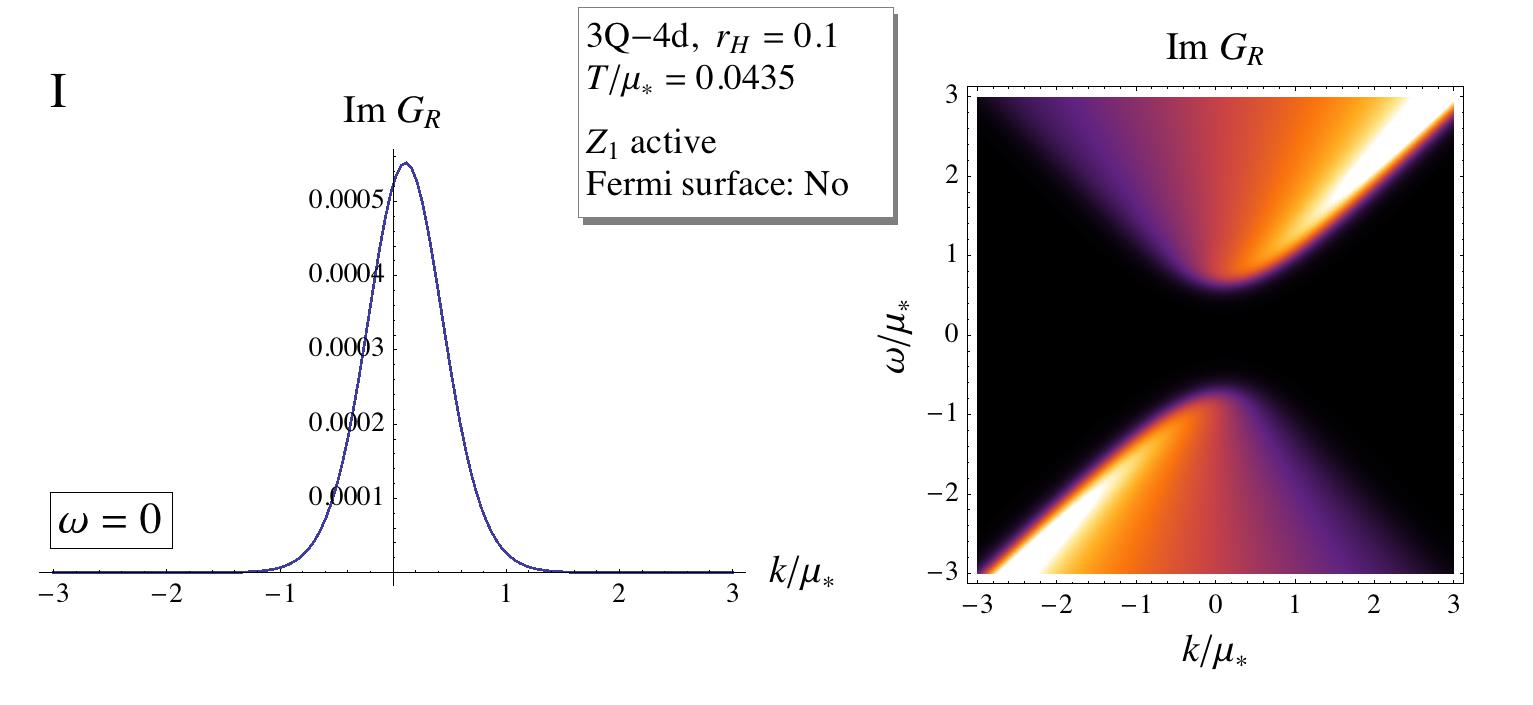}\hskip\ColumnSpacing\includegraphics[width=\PaneWidth]{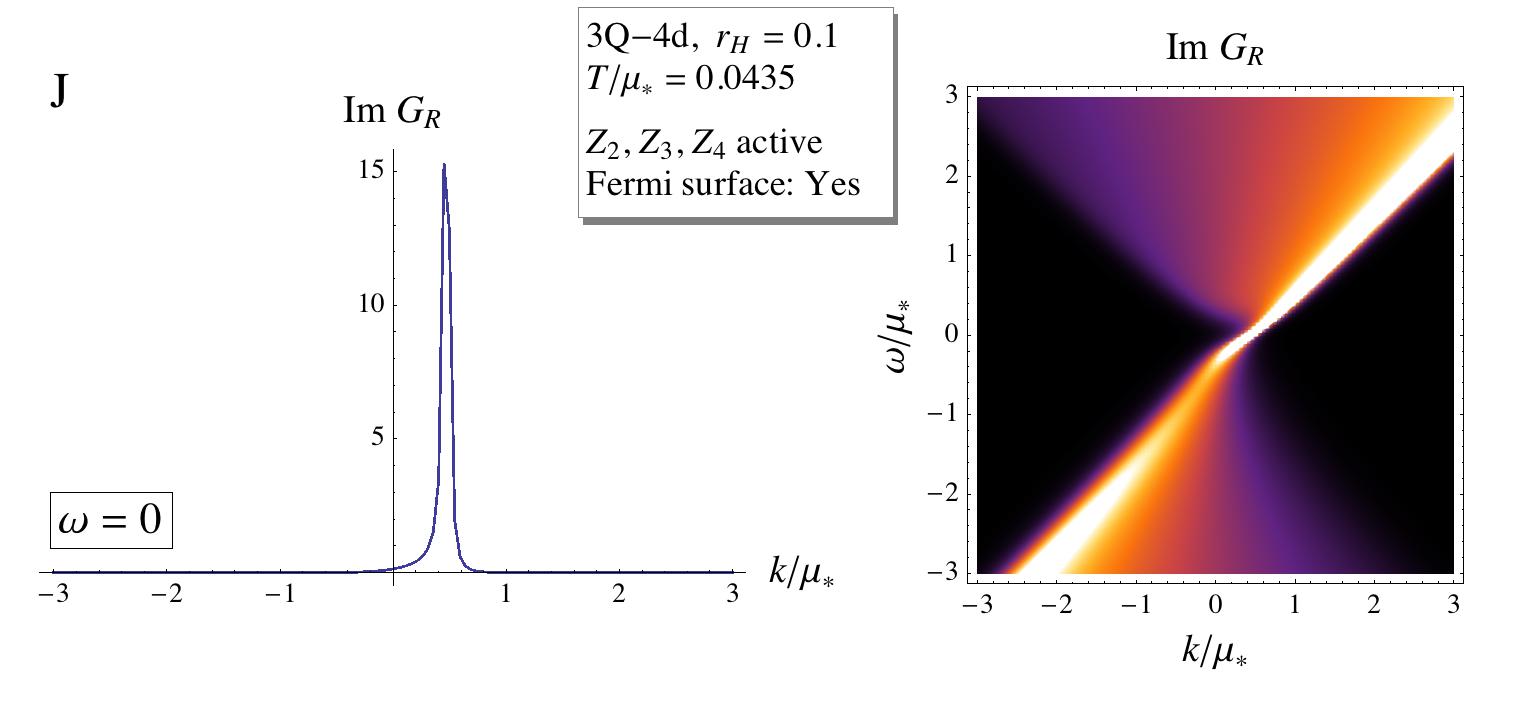}}\vskip\RowSpacing
\hbox{\includegraphics[width=\PaneWidth]{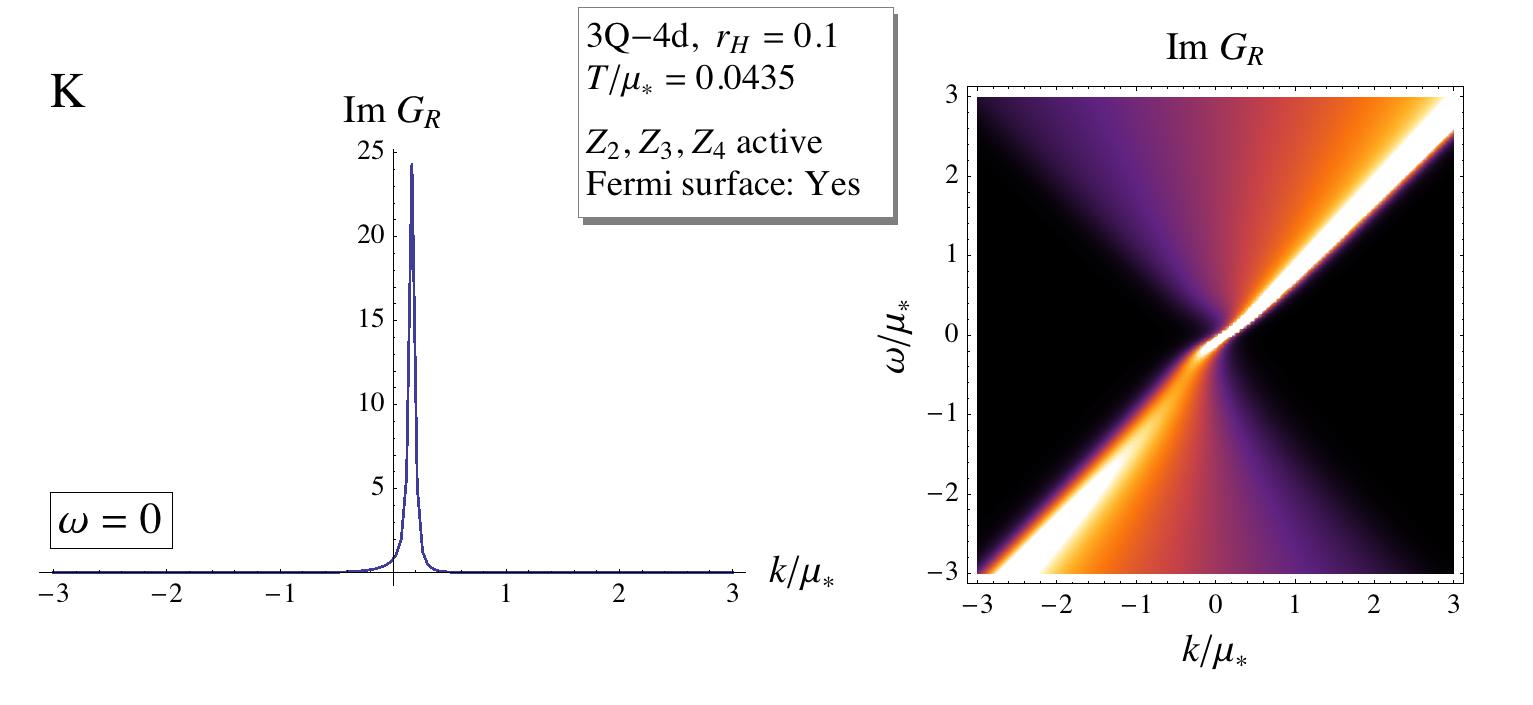}\hskip\ColumnSpacing\includegraphics[width=\PaneWidth]{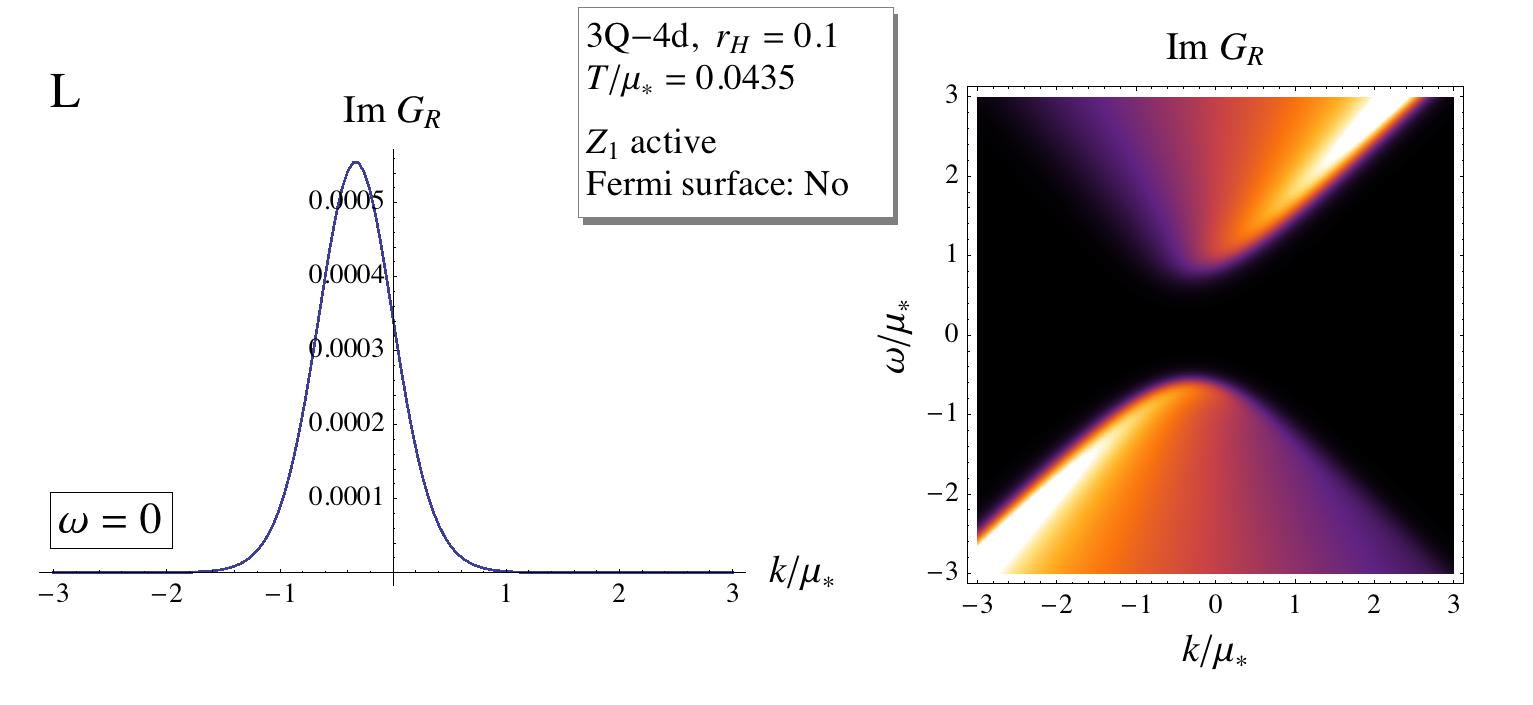}}\vskip\RowSpacing
\hbox{\includegraphics[width=\PaneWidth]{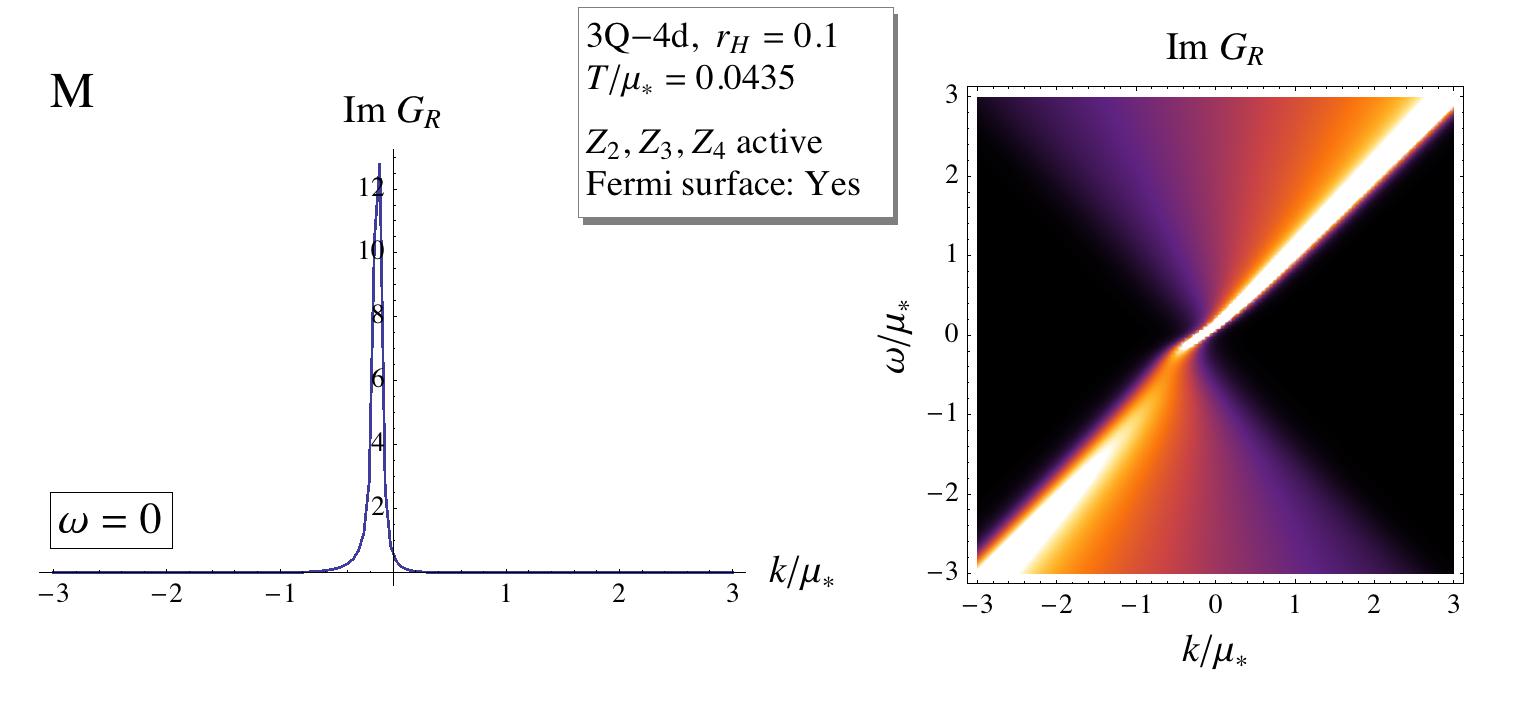}}\vskip\RowSpacing
\caption{The spectral weight of fermionic Green's functions in charged black hole backgrounds of four-dimensional gauged supergravity, continued from figure~\ref{FourDExamples}.  Accuracy of the height of the peaks in cases J, K, and M is limited by the resolution on the $k$ axis.}\label{FourDExamplesMore}
\end{figure}

Results of numerics for the spectral weight $\Im G_R(\omega,k)$ are shown in figure~\ref{FiveDExamples}-\ref{FourDExamplesMore}.  Applying the width criterion, the $\omega$ criterion, and the magnitude criterion, we judged in each case whether a holographic Fermi surface existed.  The last two columns in table~\ref{FiveDTable} and the last three columns in table~\ref{FourDTable} summarize the results.  Note that for each example background, many of the fermions behave identically: for example, in the 1Q-5d background, one is insensitive to the values of $q_2$ and $p_2$ so the fermions in the first three rows of table~\ref{FiveDTable} lead to exactly the same Green's function.  Nevertheless, with a total of eight distinct cases in five dimensions, and $13$ cases in four dimensions, we are in possession of a fairly large ``data set.''

\clearpage

\clearpage

\subsection{A boson rule}
\label{BOSON}

The obvious conclusion to draw from table~\ref{FiveDTable} is that there is a holographic Fermi surface in the 1Q-5d background precisely when the dual operator involves $Z_1$, whereas for the 2Q-5d background, there is a holographic Fermi surface precisely when the dual operators involves $Z_2$ or $Z_3$.  For the 1Q-5d background, the bulk fermions whose dual involves $Z_1$ are also the ones with the largest $q_1$ for the supergravity fermion, and from a supergravity standpoint one might prefer the size of $q_1$ as a more obvious indicator of whether there will be a Fermi surface.  But for the 2Q-5d background, one cannot use such an indicator: the fermions described in lines 1, 6, and 7 in the table all have $q_2=1$, but only the ones in lines 6 and 7 have holographic Fermi surfaces.

The boson $Z_1$ has a special role in the 1Q-5d background: $\tr |Z_1|^2 \neq 0$ for this background.  More precisely, ${\cal O}_\phi \equiv \tr\left( 2|Z_1|^2 - |Z_2|^2 - |Z_3|^2 \right)$ is the operator dual to the scalar $\phi$ that enters into \eno{DiracMostlyMinus}, and $\langle {\cal O}_\phi \rangle > 0$ can be read off from the asymptotics of $\phi$; so we conclude that $Z_1$ is non-zero.  A crucial point to note is that there is a residual $SO(2)$ symmetry which rotates the phase of $Z_1$, and it is the $U(1)$ under which the 1Q-5d background is charged.  This symmetry is unbroken in the supergravity approximation, heuristically because the D3-branes spread out in a distribution in the $Z_1$ plane which, at large $N$, is $SO(2)$ symmetric.  This intuition was made precise in \cite{Kraus:1998hv,Freedman:1999gk} for certain limits of charged black holes.  In the 2Q-5d background, the opposite situation pertains: $\langle {\cal O}_\phi \rangle < 0$, so we can conclude that $Z_2$ and $Z_3$ are non-zero.  Both of them must be non-zero because there is a residual $SO(4)$ symmetry of the two-charge background that rotates $Z_2$ into $Z_3$.  The $U(1)$ under which the 2Q-5d background is charged is part of this $SO(4)$, and it is unbroken.

We are thus led to conjecture a ``boson rule'' for the existence of holographic Fermi surfaces: {\it A supergravity fermion dual to a composite operator $\tr \lambda Z$ will show a holographic Fermi surface if and only if $Z$ (or, more precisely, an appropriate low-dimension color singlet single-trace operator built from $Z$) has a non-zero expectation value.}

The boson rule makes sense from the point of view suggested in \cite{DeWolfe:2011aa}, namely that holographic Fermi surfaces should be understood as Fermi surfaces of colored fermions in the dual field theory, not color singlets.  Simplistically, we will term this point of view the gaugino interpretation, although in a strongly interacting system without Lorentz invariance or supersymmetry, the true colored degrees of freedom could be more complicated than just gauginos.\footnote{A competing point of view (see for example \cite{Huijse:2011hp}) is the idea that the operator $\tr \lambda Z$ creates a color singlet ``mesino,'' and because mesinos are present in some density in the dual field theory, a Fermi surface is observed.  While we cannot claim to exclude the mesino interpretation based on the supergravity results presented in this paper, the success of the boson rule seems to us a significant point in favor of the gaugino interpretation.}  Acting in the maximally symmetric vacuum of ${\cal N}=4$ super-Yang-Mills theory, $\tr \lambda Z$ produces only multi-particle states, the simplest of which is a gaugino and a scalar.  Singularities in the two-point function of $\tr \lambda Z$ will have only cuts if all one can get from one insertion of $\tr \lambda Z$ is a multi-particle state with momentum shared arbitrarily between $\lambda$ and $Z$.  But if $Z$ has a condensate, then there should be a finite amplitude for $\tr \lambda Z$ to add a single $Z$ quantum to the condensate with zero momentum, injecting all the momentum available into the gaugino.  Then if there is a gaugino Fermi surface, it will show up as a singularity in the two-point function of $\tr \lambda Z$.

With the logic of the previous paragraph in mind, we should consider what a violation of the boson rule would mean.  A ``weak'' violation would be finding no Fermi surface for an operator involving a scalar that has an expectation value.  This might only signal that the color adjoint fermion that the operator in question can create happens to have no Fermi surface, or that for some more subtle reason the fermion is created in a way that fails to find the Fermi surface.  A ``strong'' violation would be finding a Fermi surface for an operator whose scalar component has no expectation value.  This would appear more troublesome for the gaugino interpretation because it would bring back the question of how a color-singlet composite operator can effectively create a colored single-particle state.

With the boson rule in mind, let's examine the $AdS_4$ cases whose study was recently made possible by the detailed supergravity analysis of \cite{DeWolfe:2014ifa}.  The punch-line is clear from table~\ref{FourDTable}: the boson rule works in every case.

\subsection{Gapped behavior}

As is clear from figure~\ref{FiveDExamples} and \ref{FourDExamplesMore}, the 2Q-5d and 3Q-4d backgrounds show very sharp peaks for some fermions, and strongly suppressed spectral weight for others.  These backgrounds are the only ones where parametrically suppressed spectral weight arises.  The suppression is associated with the gap, described in \cite{DeWolfe:2012uv,DeWolfe:2013uba,DeWolfe:2014ifa} as relating to an $AdS_3$ factor in a higher-dimensional lift of the gauged supergravity solutions.  The gapped behavior can be made more explicit by the following calculation of the spectral weight at $\omega=k=0$ for case 1F, which is the neutral fermion in the 2Q-5d background.  For this particular case, the differential equation for ${\cal I}$, ${\cal J}$, and ${\cal K}$ becomes very simple: it reads
 \eqn{SimpleIJKform}{
  \partial_r {\cal I} = -2X {\cal K} \qquad\qquad
  \partial_r {\cal J} = 0 \qquad\qquad
  \partial_r {\cal K} = -2X {\cal I} \,.
 }
Subject to the boundary conditions ${\cal I}={\cal J}=0$, ${\cal K}=1$ at the horizon, the solution to \eno{SimpleIJKform} is
 \eqn{SimpleIJKsoln}{
  {\cal I} = -\sinh e^{2 \int_{r_H}^r d\tilde{r} \, X(\tilde{r})} \qquad\qquad
  {\cal J} = 0 \qquad\qquad
  {\cal K} = \cosh e^{2 \int_{r_H}^r d\tilde{r} \, X(\tilde{r})} \,.
 }
Working through the formulas \eno{IJKFar}-\eno{GRatios}, we see that
 \eqn{XinftyRule}{
  \Im G_R(0,0) = r_H e^{-2X_\infty} \qquad\hbox{where}\qquad
   X_\infty = \int_{r_H}^\infty dr \left[ X(\tilde{r}) - {1 \over 2r} \right] \,.
 }
Working with the explicit form of $X(r)$, it is not too hard to show that $X_\infty \sim {\mu_* \over T}$, where $\sim$ indicates proportionality up to a factor of order unity and possibly logarithmic corrections.  Thus the factor $e^{-2X_\infty}$ is precisely what one expects for a thermally activated, gapped system.  No such exact analysis seems to be possible for any of the 3Q-4d cases, but a reasonable estimate of $\Im G_R(0,0)$ for cases 3I and 3L can nevertheless be given along similar lines.  The estimate is
 \eqn{XinftyEstimate}{
  \Im G_R(0,0) \sim e^{-2X_\infty} \qquad\hbox{where}\qquad
   X_\infty = \int_{r_H}^\infty dr \, X(\tilde{r}) \sim {\mu_* \over T} \,.
 }
In passing we note that case 2E admits an exact solution at $\omega=k=0$, namely $\Im G_R(0,0) = e^{-2X_\infty}$ where $X_\infty = \int_{r_H}^\infty dr \, X(\tilde{r})$.  For small $r_H$, one finds in this case that $\Im G_R(0,0) \approx {1 \over 4} r_H \approx {1 \over 4} \mu_d^2$, which cannot be described as gapped behavior.

\subsection{Toward a fermion rule}
\label{FERMION}

It's ironic that the choice of bosonic field $Z$ in the dual gauge theory operator $\tr \lambda Z$ appears to play the determining role in the presence or absence of a holographic Fermi surface.  We may well ask, what role does the choice of fermionic field $\lambda$ play?  A hint comes from comparing the two 2Q-5d cases with a Fermi surface: 1G and 1H.  The active boson is the same in each case, but the values of $k_F$ are very different: clearly non-zero for 1G, and either zero or very small for 1H.  Inspection of table~\ref{FiveDTable} suggests a plausible explanation: The gauginos involved in case 1H are neutral under the $U(1)$ carried by the 2Q-5d background, whereas the gauginos involved in case 1G are charged.  At weak coupling, we would immediately conclude that there is no Fermi sea at all for the neutral fermions, so if there are on-shell quasi-particle excitations near $\omega=0$, they should be close to $k=0$.

We turn to the 3Q-4d background to see if our explanation of the 2Q-5d results generalizes.  Unfortunately, none of the colored fermions are neutral under the $U(1)$ carried by the 3Q-4d background.  Nevertheless, it is noteworthy that for the case shown in figure~3J, $k_F$ is significantly larger than for cases 3K and 3M, and correspondingly the colored fermion for case 3J has $U(1)$ charge $3$ as opposed to charge $1$ for the colored fermions involved in cases 3K and 3M.  The values of $k_F$
are fairly small for cases 3K and 3M, but from the point of view of the gaugino interpretation, it would be surprising if $k_F \to 0$ as one passes to extremality for cases 3K and 3M: At finite chemical potential it would make more sense to have a Fermi sea of the charge $1$ colored fermions, even if it is rather smaller than the Fermi sea of charge $3$ colored fermions.

In the 1Q-5d and 2Q-4d backgrounds, the $U(1)$ chemical potential vanishes in the extremal limit while the temperature approaches a positive constant, so $\mu_*$ as defined in \eno{MuFiveDef} or \eno{MuFourDef} is $T$-dominated sufficiently close to extremality.  For the 1Q-4d background, $\mu_*$ is again $T$-dominated sufficiently close to extremality, in this case because the chemical potential vanishes faster than the temperature as one approaches extremality.  Near the extremal limit, then, it is reasonable to expect that any candidate holographic Fermi surfaces will have $k_F/\mu_* \to 0$.  This expectation is rooted less in the gaugino interpretation than in supergravity, where without a significant electric field it is harder to see why a fermionic normal mode (or a long-lived quasi-normal mode) would exist.  And it is satisfying to see in figures~\ref{FiveDExamples} and~\ref{FourDExamples} that all the Fermi surfaces in these cases have $k_F \lsim 1/2$, even though we are not all that close to the $T$-dominated limit, particularly for the 2Q-4d background.  (Numerical studies very close to an extremal limit would be aided by a more sophisticated treatment of the near-horizon limit than we have implemented.)

Within the overall trend of modest or small $k_F/\mu_*$ for the 1Q-5d, 1Q-4d, and 2Q-4d backgrounds, we can still ask when $k_F$ is suppressed by charge neutrality of the colored fermion.  The only case where this happens is the one illustrated in figure~2G, pertaining to the 2Q-4d background.  Indeed, the value of $k_F$ in this case is small---in fact, consistent with zero.  Certainly $k_F$ for case 3G is smaller than for case 3F, in the same background but with a charged colored fermion participating in the field theory operator.

A ``fermion rule'' which summarizes the phenomena observed can be formulated as follows: {\it For a supergravity fermion dual to a composite operator $\tr \lambda Z$ which exhibits a holographic Fermi surface, the value of $k_F$ is suppressed, though it may not vanish, when $\lambda$ is neutral under the $U(1)$ charge of the black hole.}  
One might hope for a more quantitative relation between $k_F$ and the dot-product of $\lambda$'s $U(1)$ charges with the $U(1)$ chemical potentials, and/or some Luttinger-style relation between $k_F$ and the charge density, but our numerical exploration to date is not sufficient to provide definite results in these directions.

\section{Discussion}

A key question in the study of fermions in charged anti-de Sitter black hole backgrounds is whether the Fermi surfaces observed should be attributed to color singlet fermions or colored fermions.  In formulating the ``boson rule'' of section~\ref{BOSON} and the ``fermion rule'' of section~\ref{FERMION}, we are trying to remain agnostic on this question and simply provide a concise summary of the results found by means of supergravity.  However, with the boson rule in hand, it is hard to remain agnostic.  It looks like a symmetry-preserving bosonic condensate pushes all the momentum from an insertion of $\tr \lambda Z$ into the colored fermion, and the singularity that arises comes from a Fermi surface of colored fermions.

One can try to go further and consider other types of operators.  Indeed, this has already been done in the literature.  For example, in \cite{Gauntlett:2011mf,Belliard:2011qq,Gauntlett:2011wm} no Fermi surface singularities were observed in the two-point function of the supercurrent.  This makes sense from the perspective of the previous paragraph, because the supercurrent includes terms which are schematically of the form $\tr \lambda \partial Z$.  Although $Z$ may have a condensate, the derivative on $Z$ means that from an insertion of the supercurrent there is a vanishing amplitude to put zero momentum into $Z$ and leave all the momentum in $\lambda$.  In \cite{DeWolfe:2012uv}, supergravity fermions dual to operators of the form $\tr F^+ \lambda$ were investigated (where $F^+$ is the self-dual part of the field strength), with the result that no Fermi surface singularities were observed.  This makes sense since composite operators built from powers of the field strength have no expectation value: for example, if $\tr F^2$ or $\tr F\tilde{F}$ had expectation values, one would expect to find a non-trivial dilaton or axion profile in the supergravity solutions, but such profiles are absent in the backgrounds under investigation.

It is worth noting that we have focused entirely on special backgrounds which have no $AdS_2$ region in the infrared.  To us this seems like a good thing, given that the $AdS_2$ regions in the extremal limit of more general black holes have non-zero entropy at zero temperature, which is hard to understand from a field theory perspective.  If we don't understand the zero point entropy in field theory, can we expect to understand the presence or absence of Fermi surfaces?  Nevertheless, let's be bold and try to apply the boson rule to black holes with $AdS_2$ near-horizon regions.  The first thing we must ask is which of the field theory scalars have expectation values.  The answer, plausibly, is all of them.  This is not unreasonable even for the special anti-de Sitter Reissner-Nordstrom solutions which have no supergravity scalar profiles, because the protected color-singlet operators, like $\tr\left( 2|Z_1|^2 - |Z_2|^2 - |Z_3|^2 \right)$, can vanish through cancellations among terms.  Thus, the naive guess based on the boson rule is that all operators of the form $\tr \lambda Z$ will have holographic Fermi surfaces.  The trouble is, that's not true \cite{DeWolfe:2012uv,DeWolfe:2014ifa}.  It is intriguing, however, that all fermions studied in \cite{DeWolfe:2014ifa} show either a pole or a zero at $\omega=0$ (or more than one such feature) for all $AdS_2$ backgrounds investigated.  Clearly, a more systematic account is desirable.

Comparisons of our calculations with zero-temperature results from \cite{DeWolfe:2013uba,DeWolfe:2014ifa} show good agreement.  For the 2Q-5d background, there is a good quantitative match between the Fermi momentum obtained from cases 1G and 1H on one hand, and the position of the pole at $\omega=0$ found in \cite{DeWolfe:2013uba} for fermions A and B on the other.  The pole for Fermion B is at small but non-zero $k_F$---an interesting point relative to the fermion rule, since this is a case where the gaugino has no charge under the $U(1)$ that is active in the 2Q-5d background.  Likewise, for the 3Q-4d background, one can successfully compare the Fermi momentum between cases 3J, 3K, and 3M on one hand, and the position of the pole at $\omega=0$ as found in \cite{DeWolfe:2014ifa} for fermions of classes 2, 1, and 3, respectively, on the other.  In other cases, where $\mu_*$ is $T$-dominated near extremality, the vanishing of $k_F/\mu_*$ as one approaches extremality makes it more difficult to find a direct quantitative comparison; however, presence or absence of poles for a given fermion matches between our analysis and those of \cite{DeWolfe:2013uba,DeWolfe:2014ifa} in all cases that we have been able to check.

\section*{Acknowledgments}

We have benefited from communications with O.~Henriksson and M.~Randeria.  We particularly thank C.~Rosen and O.~DeWolfe for detailed comments on the draft, including remarks on the match between zero-temperature and low-temperature results.  This work was supported in part by the Department of Energy under Grant No.~DE-FG02-91ER40671.  The work of C.C.H.~was also supported in part by the \'Ecole Normale Sup\'erieure.

\bibliographystyle{JHEP}
\bibliography{cases}

\end{document}